\documentclass[a4paper,twocolumn,superscriptaddress,prx, aps, floatfix ]{revtex4-2}
\usepackage{latexsym,amsmath}
\usepackage{amsbsy}
\usepackage[pdftex]{graphicx}
\usepackage{amssymb}
\usepackage{epstopdf}
\usepackage{verbatim}
\usepackage[usenames]{color}
\usepackage{capt-of} 
\usepackage{float}
\usepackage{ esint }
\usepackage{hyperref}
\usepackage{graphicx}
\usepackage[normalem]{ulem}

\usepackage{color}

\begin{document}

\title{Renormalization of networks with weak geometric coupling}

\author{Jasper van der Kolk}
\email{jasper.vanderkolk@ub.edu}
\affiliation{Departament de F\'isica de la Mat\`eria Condensada, Universitat de Barcelona, Mart\'i i Franqu\`es 1, E-08028 Barcelona, Spain}
\affiliation{Universitat de Barcelona Institute of Complex Systems (UBICS), Barcelona, Spain}

\author{Mari\'an Bogu\~n\'a}
\email{marian.boguna@ub.edu}
\affiliation{Departament de F\'isica de la Mat\`eria Condensada, Universitat de Barcelona, Mart\'i i Franqu\`es 1, E-08028 Barcelona, Spain}
\affiliation{Universitat de Barcelona Institute of Complex Systems (UBICS), Barcelona, Spain}

\author{\\M. \'Angeles \surname{Serrano}}
\email{marian.serrano@ub.edu}
\affiliation{Departament de F\'isica de la Mat\`eria Condensada, Universitat de Barcelona, Mart\'i i Franqu\`es 1, E-08028 Barcelona, Spain}
\affiliation{Universitat de Barcelona Institute of Complex Systems (UBICS), Barcelona, Spain}
\affiliation{Instituci\'o Catalana de Recerca i Estudis Avan\c{c}ats (ICREA), Passeig Llu\'is Companys 23, E-08010 Barcelona, Spain}

\begin{abstract} 
The Renormalization Group is crucial for understanding systems across scales, including complex networks. Renormalizing networks via network geometry, a framework in which their topology is based on the location of nodes in a hidden metric space, is one of the foundational approaches. However, the current methods assume that the geometric coupling is strong, neglecting weak coupling in many real networks. This paper extends renormalization to weak geometric coupling, showing that geometric information is essential to preserve self-similarity. Our results underline the importance of geometric effects on network topology even when the coupling to the underlying space is weak. 
\end{abstract}

\maketitle


The Renormalization Group remains an essential tool in statistical physics to study systems at different length scales, and for revealing the scale invariance and universal properties of critical phenomena near continuous phase transitions where fluctuations are strong~\cite{Tauber2012}. The simplest technique for processes on regular lattices is that of the block spin method proposed by Kadanoff~\cite{Kadanoff1966}, where blocks of nearby nodes are grouped together into supernodes whose state is determined by some averaging rule. Extending this method to complex networks is complicated by their small world property, which makes the concept of closeness fuzzy and hinders the definition of supernodes~\cite{Watts1998}.

In complex networks, different renormalization schemes have been proposed. Some are based on ensemble self-similarity~\cite{garuccio2023multiscale}, while others use proximity measures. In the box covering method, nodes are grouped depending on topological distance~\cite{Song2005}, and in Laplacian renormalization closeness is based on diffusive distance~\cite{Villegas2023}. As an alternative, network geometry~\cite{Boguna2021,serrano_boguna_2022} is based on the assumption that nodes lie in an underlying metric space such that closer nodes are more similar and therefore more likely to be connected, and so it offers a natural framework for describing and renormalizing networks. The family of models with latent hyperbolic geometry have shown to be highly effective in generating network structures with realistic topological features~\cite{Serrano2008,krioukov2010hyperbolic,gugelmann2012random,candellero2016clustering,Fountoulakis2021,abdullah2017typical,friedrich2018diameter,muller2019diameter}, percolation characteristics~\cite{serrano2011percolation,fountoulakis2018law}, spectral aspects~\cite{kiwi2018spectral}, and self-similarity~\cite{Serrano2008}.

These geometric models have served as a foundation for defining a renormalization group for complex networks~\cite{GarciaPerez2018,Zhenge2021}, in which adjacent nodes are coarse-grained into supernodes on the basis of their coordinates in their latent geometry. Building upon this concept, the geometric renormalization (GR) approach has revealed that scale invariance is a pervasive symmetry in real networks~\cite{GarciaPerez2018}. From a practical perspective, GR has also enabled the generation of scaled-down self-similar replicas---an essential tool for facilitating the computationally challenging analysis of large networks. Additionally, when combined with scaled-up replicas produced through a fine-graining reverse renormalization technique~\cite{Zhenge2021}, it provides a means to explore size-dependent phenomena.

GR typically assumes that real networks, which display high levels of clustering, are strongly coupled to their latent geometry. However, the model presents a phase transition at a certain critical coupling of the geometry and the topology of a network where it goes from a strongly geometric regime with a finite density of triangles in the thermodynamic limit to a weakly geometric regime where this quantity vanishes~\cite{Serrano2008}. It has been shown that many real networks with significant triangle densities~\cite{vanderkolk2024} are better described in a quasi-geometric domain of the weakly geometric region, in which the decay of the clustering coefficient is very slow~\cite{vanderKolk2022}. In this paper, we develop GR in the regime of weak geometric coupling and apply the extended renormalization scheme to a set of real networks in the quasi-geometric domain. We show that, in this regime, geometric information is essential for obtaining self-similarity in important network measures across scales.

The geometric renormalization group for complex networks introduced in Ref.~\cite{GarciaPerez2018} is constructed upon the $\mathbb{S}^1$ model~\cite{Serrano2008}. Specifically, the connectivity in the $\mathbb{S}^1$ model is determined by `popularity', which is related to the degree of a node, and by `similarity', which encodes for all other inherent properties of the nodes. The similarity dimension is explicit; nodes are placed on a circle and given angular coordinates $\theta_i$. The circle has radius $R=N/(2\pi)$, such that the density of points remains constant for different network sizes. In contrast, the popularity dimension is encoded by a hidden degree $\kappa_i$ drawn from some arbitrary distribution $\rho(\kappa)$, typically a power law with exponent $\gamma$. Two nodes are connected with probability  

\begin{equation}
	p_{ij} = \frac{1}{1+\frac{(R\Delta\theta_{ij})^ \beta}{(\hat{\mu}\kappa_i\kappa_j)^{\max{(1,\beta)}}}}
	\label{eq:ConProb},
\end{equation}
where $\hat{\mu}$ sets the average degree $\langle k\rangle$ and where $\Delta\theta_{ij}=\pi-|\pi-|\theta_i-\theta_j||$ defines the angular distance between the two nodes. The parameter $\beta$, often referred to as the inverse temperature in analogy to statistical physics, sets the level of clustering in the network. The $\mathbb{S}^1$ model has realizations that are realistic and maximally random, in that their connection probability defines a ensemble that maximizes entropy~\cite{Boguna2020}. Note that the $\mathbb{S}^1$ model has an isomorphic equivalent in the hyperbolic plane, the $\mathbb{H}^2$ model~\cite{krioukov2010hyperbolic}, in which the popularity dimension is made explicitly geometric by mapping the hidden degree $\kappa$ to a radial coordinate $r$. 

The inverse temperature $\beta$ calibrates the coupling between the underlying metric space and the geometry: when $\beta$ is high, $p_{ij}$ is large only when $\Delta\theta_{ij}$ is small or $\kappa$s are large, which implies that the network contains mostly short ranged links. Conversely, when $\beta\rightarrow0$, the dependence of the connection probability on the angular coordinate is lost and long range links are as likely as short range ones. Note that the dependence on the popularity dimension does not vanish, and that at $\beta=0$ our ensemble is equivalent to that of the hyper-soft configuration model~\cite{vanderHoorn2018}. It has been shown that, in the thermodynamic limit, clustering is finite for $\beta>1$ and vanishes when $0\leq\beta\leq1$~\cite{Serrano2008}. However, the slow approach to the thermodynamic limit in the region $0.5\lesssim\beta< 1$ below the transition implies that certain real networks that show significant clustering can be better described in this so-called quasi-geometric domain~\cite{vanderKolk2022}. 

The first step in GR is to define non-overlapping sectors along the $\mathbb{S}^1$ circle containing each $r$ consecutive nodes. To determine the coordinates of the nodes in the geometric space and, hence, which nodes are consecutive in the similarity space, real networks can be embedded in the $\mathbb{S}^1/\mathbb{H}^2$-model by finding the coordinates that are most congruent with the network topology. 
One embedding tool is Mercator~\cite{GarciaPerez2019}, which was recently extended to handle weakly geometric networks~\cite{vanderkolk2024} and higher dimension metric spaces~\cite{jankowski2023d}. The second step is to coarse grain the nodes within a group to form a single supernode, whose angular coordinate and hidden degree are functions of the coordinates and hidden degrees of its constituents. It is essential that the supernode order along the circle preserves the order of nodes in the original layer. The connectivity of the new network is defined by connecting two supernodes if any pair of their respective constituents are connected. This procedure can be repeated iteratively starting from the original layer $l=0$. Each layer $l$ is then $r^l$ times smaller than the original network. This then defines the renormalization group flow.

In the following, we extend this procedure to the region $\beta\leq1$. We use a compact notation that includes the results in~\cite{GarciaPerez2018} for $\beta>1$ (see Supplementary Information I for details~\cite{supp}). Demanding that the connection probability remains invariant under the renormalization flow independently of $\beta$, i.e. that each scaled-down network is congruent with the $\mathbb{S}^1$-model for weak or strong geometric coupling, leads to the transformation
\begin{equation}
	\kappa_\sigma^{(l+1)} = \left[\sum_{i\in \mathcal{S}(\sigma)}\left(\kappa_i^{(l)}\right)^{\max(1,\beta)}\right]^{\frac{1}{\max(1,\beta)}}
	\label{eq:evolution_kappas}
\end{equation}
for the hidden degrees. Here, $\mathcal{S}(\sigma)$ represents the set of constituent nodes of supernode $\sigma$. Note that in the weak coupling regime $\max(1,\beta)=1$, reducing the definition to a simple sum. This definition satisfies the semi-group property,  i.e. renormalizing twice with groups of $r$ is the same as renormalizing once with groups of $r^2$. The global parameters flow as $R^{(l+1)}=R^{(l)}/r^l$, $\hat\mu^{(l+1)}=\hat\mu^{(l)}/r^{\min(1,\beta)}$, and $\beta^{(l+1)}=\beta^{(l)}$.

The flow of the average hidden degree can be derived from the flow of the hidden degrees. In the region $\beta\leq1$ (the case $\beta>1$ was already investigated in Ref.~\cite{GarciaPerez2018}) the hidden degree of a supernode is simply given by the sum of the hidden degrees of its constituents, which implies
\begin{equation}
	\langle \kappa_\sigma^{(l+1)}\rangle=\Big\langle\sum_{i\in \mathcal{S}(\sigma)}\kappa^{(l)}_i\Big\rangle=\sum_{i\in \mathcal{S}(\alpha)}\langle\kappa^{(l)}_i\rangle=r\langle\kappa^{(l)}\rangle\label{eq:evolution_average_kappa}.
\end{equation}
Thus, $\langle \kappa ^{(l+1)} \rangle = r^{\xi}\langle \kappa^{(l)}\rangle$ where $\xi = 1$ for $\beta\leq 1$. Using this, it can be shown that the flow of the average degree is $\langle k ^{(l+1)} \rangle = r^{\nu}\langle k^{(l)}\rangle$, where $\nu=2\xi-1=1$. The flow of the average degree in the weakly geometric regime is, thus, inversely proportional to the flow of the system size, which means that the amount of links $E=N\langle k\rangle/2$ is a constant under renormalization, i.e. that no links are lost as one performs GR steps. This result implies that, on average, there is only one connection between the constituents of a pair of supernodes. This has to do with the fact that, for $\beta\leq 1$, connections are long ranged in the thermodynamic limit. It is therefore exceedingly unlikely that a node is connected to two nodes so close together in the latent space that when we perform a GR step they end up in the same supernode.

Concerning the flow of the angular coordinate, any transformation that preserves the order of nodes in the original layer would work as long as it preserves the rotational invariance of the original model. We choose
\begin{equation}
	\theta_\sigma^{(l+1)} = \frac{\sum_{i\in\mathcal{S}(\sigma)}\left(\kappa_i^{(l)}\right)^{\max(1,\beta)}\theta_i^{(l)}}{\sum_{i\in\mathcal{S}(\sigma)}\left(\kappa_i^{(l)}\right)^{\max(1,\beta)}},\label{eq:evolution_thetas}
\end{equation}
the weighted average of the constituent nodes. This is a slightly different definition as the one used in Ref.~\cite{GarciaPerez2018}, where the weighted power mean with exponent $\max(1,\beta)$ was used. However, for larger $\beta$ this introduces a bias towards larger constituent $\theta$'s, which is not in line with the rotational symmetry of the system. The definition in Eq.~\eqref{eq:evolution_thetas} satisfies the semi-group property. 

\begin{figure}[t]
	\centering
	\includegraphics[width=1\columnwidth]{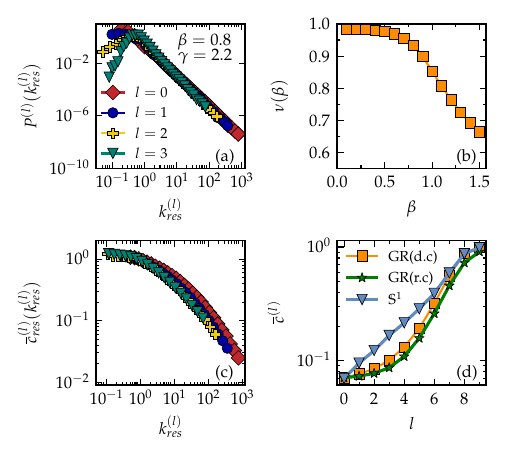}
	\vspace{-0.8cm}
	\caption{\textbf{(a)} The log-binned degree distribution $P^{(l)}(k^{(l)})$ as a function of the rescaled degrees $k_{res}^{(l)}=k/\langle k^{(l)}\rangle$. \textbf{(b)} The exponent $\nu$ in $\langle k^{(l+1)}\rangle=r^{\nu}\langle k^{(l)}\rangle$, as a function of $\beta$. \textbf{(c)} Rescaled average local clustering per degree class $\overline{c}^{(l)}_{res}(k^{(l)})=(\overline c^{(l)}(k^{(l)}))/\overline c^{(l)}$ as a function of the rescaled degrees. \textbf{(d)} Average local clustering coefficient $\overline c^{(l)}$ as a function of the layer $(l)$. We display the flow under standard GR with deterministic links (orange squares), GR where links are made probabilistically (green stars), and new independent $\mathbb{S}^1$ realizations created in every layer (blue triangles). In this latter case, the networks size and the average degree match the GR in every layer. The original networks were generated with the $\mathbb{S}^1$ model for $N=65536$ and $\langle k\rangle=6$. }	\label{fig:RG_1}
\end{figure}
In Fig.~\ref{fig:RG_1}, we show the behavior of several network properties in the flow of synthetic scale-free networks generated with the $\mathbb{S}^1$ model. In Fig.~\ref{fig:RG_1}a, the tail of the complementary cumulative degree distribution of rescaled degrees $k_{res}^{(l)}=k^{(l)}/\langle k^{(l)}\rangle$ for $\beta=0.8$ in the quasi-geometric domain is self-similar under renormalization. This self-similarity is also proven analytically in Supplementary Information II~\cite{supp}. For large enough $l$, this self-similarity will always be lost for finite systems like real networks. This is because the finite size induces a cut-off in the degree distribution, rendering its variance finite and therefore leading to the applicability of the central limit theorem, resulting in a Gaussian distribution.

In Fig.~\ref{fig:RG_1}b, we plot the dependence of the exponent $\nu$ characterizing the flow of the average degree as a function of $\beta$. 
As discussed above, in the region $\beta\leq1$ no edges get destroyed in the renormalization flow as the long range nature of links in this regime makes it extremely unlikely that two or more edges connect nodes in the same two supernodes. 
However, such situations do arise for finite systems, leading to the loss of links along the flow and, thus, to $\nu<1$, as can be observed in Fig.~\ref{fig:RG_1}b. This finite size effect is stronger the closer to $\beta=1$, and can therefore be seen as quasi-geometric behavior. When $\beta>1$, the exponent $\nu$ decreases even further and we enter in the geometric regime described in Ref.~\cite{GarciaPerez2018}. 

In Fig.~\ref{fig:RG_1}c we display the average local clustering coefficients per degree class, which is again self-similar when rescaled as $\overline{c}^{(l)}_{res}(k^{(l)})=(\overline c^{(l)}(k^{(l)}))/\overline c^{(l)}$, where $\overline c^{(l)}$ is the average local clustering coefficient. Rescaling is necessary because $\overline c^{(l)}$ is not conserved under the RG flow for $\beta<1$. This is confirmed by Fig.~\ref{fig:RG_1}d, where the orange squares represent the evolution of $\overline c^{(l)}$ as a function of the renormalization step $l$ for networks at $\beta=0.8$. This behavior is in contrast to the situation for $\beta>1$, where $\overline c$ only depends on the inverse temperature $\beta$, which is unaffected by the renormalization procedure. For $\beta\leq1$, clustering depends on the systems size and the average degree~\cite{vanderKolk2022}, which do change under the RG flow. 

This result on its own is not in tension with the notion of self-similarity as, granted the network is well described by the $\mathbb{S}^1$-model, a smaller version of a certain network should indeed have a higher clustering coefficient. However, comparing networks obtained through GR (orange squares) and with the $\mathbb{S}^1$-model (blue triangles) in Fig.~\ref{fig:RG_1}d, we see that the flows of $\overline{c}^{(l)}$ do not match. This discrepancy is caused by the fact that the largest contribution to the average local clustering coefficient comes from nodes with small degrees for which self-similarity is not fulfilled, as can be seen in Fig.~\ref{fig:RG_1}a. To prove that the discrepancy does not stem from a lack of congruence with the $\mathbb{S}^1$ connection probability, we repeated the same analysis for networks where the hidden degrees of the supernodes were generated using Eqs.~\eqref{eq:evolution_kappas} and \eqref{eq:evolution_thetas} but where the connections were made randomly following Eq.~\eqref{eq:ConProb}. In Fig.~\ref{fig:RG_1}d, this case is represented by green stars and coincides with the GR flow. The discrepancy, thus, originates in the lack of self-similarity of the hidden degree distribution at small $\kappa$.   

Geometry is still important for renormalizing networks with weak geometric coupling. To prove this, we compared GR on synthetic networks with $\beta\leq1$ with a second scheme that is explicitly non-geometric, where supernodes are created by choosing constituent nodes at random. In this scheme, the angular coordinate of the supernodes is meaningless by construction. Nevertheless, for convenience, we redefined it such that it represents a proper average even for constituent nodes that lie far away from each other. To this end we use the weighted circular mean, which takes coordinates along the circle to represent roots of unity, which are then summed, weighted by the hidden degrees of nodes in the same way as in Eq.~\eqref{eq:evolution_thetas}. Finally, the argument of the result is taken to obtain the angular coordinate of the supernode. Note that this definition reduces to Eq.~\eqref{eq:evolution_thetas} when the angular spread is small~\cite{supp}. In general, the random scheme will not lead to a conserved connection probability as the proof in Supplementary Information I~\cite{supp} requires the angular separation of nodes within a supernode to be much smaller than the angular separation between supernodes. However, one might argue that the angular coordinate is irrelevant as the regime $\beta<1$ is \textit{a priori} non-geometric in the thermodynamic limit and self-similar network copies could, thus, still be obtainable. As we show below, for finite networks this is only the case for extremely small values of $\beta\lesssim0.5$, which we call the non-geometric regime. 

\begin{figure}[h]
	\centering
	\includegraphics[width=1\columnwidth]{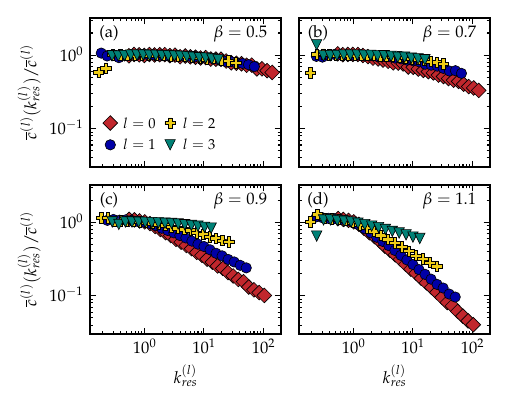}
	\vspace{-0.8cm}
	\caption{The flow of the rescaled average local clustering coefficient per rescaled degree class under the randomized coarse-graining scheme for different $\beta's$: \textbf{(a)} $\beta=0.5$, \textbf{(b)} $\beta=0.7$, \textbf{(c)} $\beta=0.9$, \textbf{(d)} $\beta=1.1$. Here, $l=0$ represents the original network and we perform three consecutive renormalization steps with $r=2$, leading to the cases $l=1,2$ and $3$. The network parameters used to generate the original networks are $\{N,\gamma,\langle k\rangle\}=\{65536,2.9,6\}$.}
	\label{fig:RG_2}
\end{figure}
We first study self-similarity of the clustering spectrum as clustering is the key property of geometric graphs due to its relation to the triangle inequality. In Fig.~\ref{fig:RG_1}c, it was shown that, for a scale-free synthetic network with $\beta=0.8$, the standard GR reveals self-similar behavior in the renormalization flow. In contrast, in Fig.~\ref{fig:RG_2}, we show the results for the randomized coarse-graining scheme. We see that self-similarity is obtained for the smallest $\beta$'s, implying that geometric information is not important here. However, the overlap between the different curves gets progressively worse as $\beta$ increases, reflecting the growing importance of the geometry. The self-similarity is lost at $\beta\approx0.7$, very close to the theoretical transition point $\beta_c'=2/\gamma$ between the non- and quasi-geometric regimes~\cite{vanderKolk2022}. The curves flatten out with $l$, implying that more and more of the clustering in the network is due to high degree nodes. This is to be expected, as the random coarse-graining scheme destroys the coupling of the network to the geometry. This leads to networks that are similar to those generated with the configuration model, where we know that most of the clustering is due to high degree nodes~\cite{Colomer-de-Simon:2012vr}.   

\begin{figure}[h]
	\centering
	\includegraphics[width=1\columnwidth]{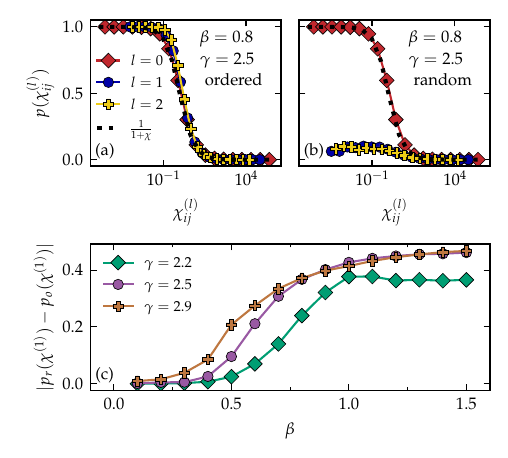}
	\vspace{-0.8cm}
	\caption{\textbf{(a,b)} The flow of the connection probability as a function of $\chi=x_{nm}^\beta/(\hat \mu\kappa_n\kappa_m)^{\max(1,\beta)}$ under the RGN where nodes are combined sequentially \textbf{(a)} or randomly \textbf{(b)} along the circle. The dotted lines give the theoretical curve. \textbf{(c)} The mean difference between the two previous cases for $l=1$ and for three different $\gamma$'s. In all cases the networks were generated with the $\mathbb{S}^1$ model with $N=65536$ and $\langle k\rangle=6$. }
	\label{fig:RG_3}
\end{figure}

To quantify further how much poorer the results of the randomized coarse-graining scheme are in comparison to GR, we measured how well the empirical connection probability of the renormalized network fits the theoretical one in the $\mathbb{S}^1$-model. After obtaining the hidden coordinates, the parameter $\chi_{ij}=x_{ij}^\beta/(\hat{\mu}\kappa_i\kappa_j)^{\max(1,\beta)}$ were determined for each pair of nodes. These values were binned logarithmically, and for each bin the proportion of links versus non-links was calculated to produce the inferred connection probability $p(\chi)$. The results of this analysis are shown in Fig.~\ref{fig:RG_3} where we have used networks in the quasi-geometric regime with $\beta=0.8$. Fig.~\ref{fig:RG_3}a shows the inferred connection probability of the different renormalized layers for the standard GR, where geometric information is used to define the supernodes. In Fig.~\ref{fig:RG_3}b, we see the same results but for the case where the nodes are chosen at random. Clearly, while GR produces self-similar copies congruent with the $\mathbb{S}^1$ connection probability, the random procedure does not. This confirms that in the quasi-geometric regime geometric information is important even though the geometric coupling is weak. 

\begin{figure*}[t!]
	\centering
	\includegraphics[width=1\textwidth]{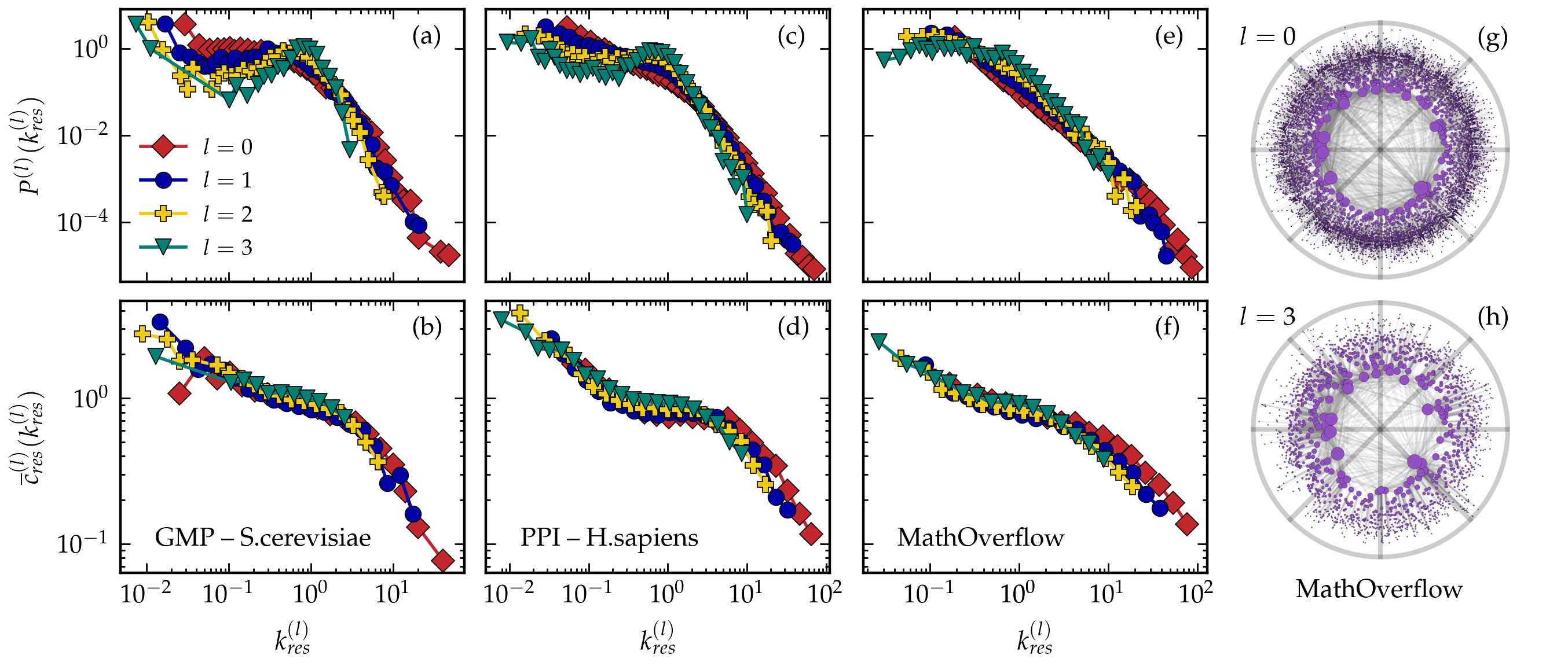}
	\caption{The degree distribution and rescaled average local clustering coefficient as functions of the rescaled degree for \textbf{(a,b)} the genetic multiplex of the yeast S.cerevisiae, \textbf{(c,d)} the Human protein-protein interaction network and \textbf{(e,f)} the interaction network of users on the online Q\&A site MathOverflow. For this last network, the $\mathbb{H}^2$ representations of the \textbf{(g)} original ($l=0$)  and \textbf{(h)} renormalized ($l=3$) networks are shown. The details of these networks are given in the SM~\cite{supp}. }   
	\label{fig:RG_4}
\end{figure*}

We plot the average difference between the connection probabilities of the two schemes at layer $l=1$ as a function of the inverse temperature $\beta$ in Fig.~\ref{fig:RG_3}c. To compute this difference, one first samples parameters $\chi_{ij}^{(l=1)}$ logarithmically. For each of these values, one finds the observed connection probability for the two schemes. One then takes the difference between these cases and averages this over the sampled distances. Once again, three different behaviors can be observed. In the geometric regime ($\beta>1$), the difference between the two methods is large. For $\beta$'s in the quasi-geometric regime, the difference decreases, and it goes to zero in the non-geometric regime. The transition point between the non- and quasi-geometric regimes shifts to higher betas when the heterogeneity of the network is increased, in line with the theoretical prediction that this transition occurs at $\beta_c'=2/\gamma$~\cite{vanderKolk2022}. The discrepancy between the curves at $\beta>1$ comes from the fact that not only similarity but also popularity plays a role in the connection probability. As this second type of information is used equivalently in the renormalization procedure regardless of how the angular coordinates are chosen, the difference between these two methods can thus be expected to be smaller when popularity dimensions plays a more important role, which is the case when the degree distribution is more heterogeneous, i.e. when $\gamma$ is smaller.  

Now that we have set up the renormalization procedure $\beta<1$ and shown that geometric information is relevant in this regime, we are able to study the self-similarity of real networks that are best described as being weakly geometric~\cite{vanderkolk2024}. In Fig.~\ref{fig:RG_4}e-f the degree distribution and clustering spectrum of several of those real networks and their scaled-down replicas are shown. In particular, we study the genetic multiplex of the nematode worm \textit{C. Elegans} (Fig.~\ref{fig:RG_4}a,b)~\cite{DeDomenico2015}, the human protein-protein interaction network (Fig.~\ref{fig:RG_4}c,d)~\cite{Hu2018} and the interaction network of users on the online Q\&A site \textit{MathOverflow} (Fig.~\ref{fig:RG_4}e,f)~\cite{Paranjape2017}. The embeddings of these networks in the quasi-geometric domain were produce with Mercator~\cite{vanderkolk2024,GarciaPerez2019}. Further details about the networks can be found in Supplementary Information III~\cite{supp}.

In all cases, the curves remain invariant under repeated application of GR. Only for large $l$ does the degree distribution tend to a more homogeneous distribution. This is once again a finite size effect. For the MathOverflow network, we show the $\mathbb{H}^2$ representation of the original (Fig.~\ref{fig:RG_4}g) and scaled-down (Fig.~\ref{fig:RG_4}h) networks. To obtain the scaled-down replica, GR with $r=2$ was performed thrice, such that the replica was $2^3=8$ times smaller than the original. We report similar results for a wide range of other networks in the Supplementary Information IV~\cite{supp}.

Finally, the embedding of a real network produces a non-homogeneous placement of nodes in the similarity space. The GR procedure inherently preserves the distribution of points along the circle as the angular coordinate of a supernode is just the weighted average of the coordinates of its constituents. This then automatically implies that the soft, and therefore structural, community structure is preserved even in the weakly geometric regimes. We can actually see an example of this in the Supplementary Information. In Fig.~S7, the distribution of nodes is clearly not homogeneous, with several regions divided by large gaps. We see that as one performs GR, these regions persist, thus preserving the soft community structure.

In summary, we have extended the geometric renormalization scheme to networks in the weakly geometric regime. The differences between the weakly geometric and the strongly geometric regions are quantitative. The different connection probabilities on both sides of the clustering transition at $\beta_c=1$ imply an altered transformation law for the hidden degrees and a different scaling of the average degree as successive renormalization steps are performed. However, these differences do not change the paradigm that the geometric renormalization scheme can produce self-similar network replicas on both sides of the transition. In fact, geometric information is essential for achieving this goal when $\beta\gtrsim0.5$. In the weakly geometric regime, self-similarity still refers to important network properties, including not only the degree distribution but also the clustering spectrum. As in the geometric case, these replicas have many applications~\cite{GarciaPerez2018}. For instance, they enable finite size scaling studies of real networks from single snapshots, exploring large phase spaces more efficiently, and the identification of communities by leveraging the mesoscopic information encoded in the different multiscale layers. In the quasi-geometric domain $0.5\lesssim\beta\leq1$, one must define supernodes by grouping consecutive nodes along the $\mathbb{S}^1$ circle in order to obtain self-similarity in the clustering spectrum and in the connection probability. This underlines the importance of geometric information for understanding the network topology even when the geometric coupling is weak. In constrast, for $\beta\lesssim0.5$ it does not matter how nodes are grouped. This implies that, here, 
the connectivity is solely determined by the degree-distribution, making them effectively non-geometric. Finally, we reveal the scale-invariance of many real networks identified in Ref.~\cite{vanderkolk2024} as living in the quasi-geometric domain of the weak coupling regime, which can be effectively renormalized using the extended GR scheme. These results prove once again the importance of the geometric renormalization approach to reveal hidden symmetries in real networks. 

The code of the Mercator embedding tool used in this paper is publicly available at \url{https://github.com/networkgeometry/mercator}.

This work was supported by grant TED2021-129791B-I00 funded by MICIU/AEI/10.13039/501100011033 and the ``European Union NextGenerationEU/PRTR''; grant PID2022-137505NB-C22 funded by MICIU/AEI/10.13039/501100011033 and by ERDF/EU; Generalitat de Catalunya grant number 2021SGR00856. M. B. acknowledges the ICREA Academia award, funded by the Generalitat de Catalunya. J.~vd~K. acknowledges support from the Ministry of Universities of Spain in the form of the FPU predoctoral contract.

\bibliography{bibliography}

\end{document}


\title{Supplemental Material for \\Renormalization of networks with weak geometric coupling}

\author{Jasper van der Kolk}
\email{jasper.vanderkolk@ub.edu}
\affiliation{Departament de F\'isica de la Mat\`eria Condensada, Universitat de Barcelona, Mart\'i i Franqu\`es 1, E-08028 Barcelona, Spain}
\affiliation{Universitat de Barcelona Institute of Complex Systems (UBICS), Barcelona, Spain}

\author{Mari\'an Bogu\~n\'a}
\email{marian.boguna@ub.edu}
\affiliation{Departament de F\'isica de la Mat\`eria Condensada, Universitat de Barcelona, Mart\'i i Franqu\`es 1, E-08028 Barcelona, Spain}
\affiliation{Universitat de Barcelona Institute of Complex Systems (UBICS), Barcelona, Spain}

\author{\\M. \'Angeles \surname{Serrano}}
\email{marian.serrano@ub.edu}
\affiliation{Departament de F\'isica de la Mat\`eria Condensada, Universitat de Barcelona, Mart\'i i Franqu\`es 1, E-08028 Barcelona, Spain}
\affiliation{Universitat de Barcelona Institute of Complex Systems (UBICS), Barcelona, Spain}
\affiliation{Instituci\'o Catalana de Recerca i Estudis Avan\c{c}ats (ICREA), Passeig Llu\'is Companys 23, E-08010 Barcelona, Spain}

\maketitle

\tableofcontents

\section{Self-similarity of the connection probability}\label{sec:self-similarity of the connection probability}
In this section we show that the connection probability 
\begin{equation}
	p_{ij}=\left(1+\frac{(R\Delta\theta_{ij})^\beta}{(\hat\mu\kappa_i\kappa_j)^{\max{(1,\beta)}}}\right)^{-1}
\end{equation}
is self-similar under renormalization if certain choices are made. In the renormalization procedure described above, supernodes in layer $l+1$ are formed by combining $r$ adjacent nodes from layer $l$. If any constituent of supernode $\sigma$, denoted by the set $\mathcal{S}(\sigma)$ is connected to any of the constituents of supernode $\tau$ they are said to be connected. The probability of this being the case is given by
\begin{equation}
	p_{\sigma\tau}^{(l+1)}=1-\prod_{(i,j)\in \mathcal{P}(\sigma,\tau)}(1-p^{(l)}_{ij}),
\end{equation}
i.e. one minus the probability that none of the constituents are connected. Here we have defined $\mathcal{P}(\sigma,\tau)=S(\sigma)\times S(\tau)$. Using that $p^{(l)}_{ij}=1/(1-x^{(l)}_{ij})$ we can rewrite this expression as
\begin{alignat}{6}
	p_{\sigma\tau}^{(l+1)}=1-\frac{1}{\prod_{(i,j)\in \mathcal{P}(\sigma,\tau)}(1+(x_{ij}^{(l)})^{-1})}\label{eq:cp_renorm1}.
\end{alignat}
The denominator of the second term can be expanded as 
\begin{widetext}
	\begin{alignat}{6}
		&\prod_{(i,j)\in \mathcal{P}(\sigma,\tau)}(1+(x^{(l)}_{ij})^{-1}) = 1 + \sum_{(i,j)\in \mathcal{P}(\sigma,\tau)}(x^{(l)}_{ij})^{-1}
		+\sum_{(i,j)\in \mathcal{P}(\sigma,\tau)}(x^{(l)}_{ij})^{-1}\sum_{(s,t)\in \mathcal{P}(\sigma,\tau)\setminus(n,m)}(x^{(l)}_{st})^{-1}+...
	\end{alignat}
\end{widetext}
We know that $x^{(l)}_{ij}=(R^{(l)}\Delta\theta^{(l)}_{ij})^{\beta^{(l)}}/(\hat{\mu}^{(l)}\kappa^{(l)}_i\kappa^{(l)}_j)^{\max(1,\beta^{(l)})}$ which is proportional to $(N^{(l)})^{\max(1,\beta^{(l)})}\gg1$. Thus, we can truncate the expansion at first order. We assume that $\Delta\theta^{(l)}_{ij}\approx\Delta\theta_{\sigma\tau}^{(l+1)}$, the distance between the two supernodes. This is because the distances between the nodes within a single supernode is generally much smaller than the distance between nodes in different supernodes. This allows us to rewrite Eq.~\eqref{eq:cp_renorm1} as 
\begin{equation}
	p_{\sigma\tau}^{(l+1)}=\left(1+\frac{\left(R^{(l)}\Delta\theta^{(l+1)}_{\sigma\tau}\right)^{\beta^{(l)}}}{\sum_{(i,j)\in \mathcal{P}(\sigma,\tau)}\left(\hat\mu^{(l)}\kappa_i^{(l)}\kappa_j^{(l)}\right)^{\max(1,\beta^{(l)})}}\right)^{-1}.\label{eq:renormalized_connection_probability}
\end{equation}
In order for this to be a proper connection probability in the renormalized layer, taking into account that $R^{(l+1)}=R^{(l)}/r$ and $\beta^{(l+1)}=\beta^{(l)}\equiv \beta$, we must demand $\hat\mu^{(l+1)}=\hat\mu^{(l)}/ r^{\min(1,\beta)}$. Furthermore, the evolution of the hidden degrees is as follows
\begin{equation}
	\kappa_\sigma^{(l+1)} = \left(\sum_{i\in \mathcal{S}(\sigma)}\left(\kappa_i^{(l)}\right)^{\max(1,\beta)}\right)^{1/\max(1,\beta)}\label{eq:kappaevol}.
\end{equation}
This transformation respects the semi-group property of the renormalization as 
\begin{alignat}{6}
	\kappa_\sigma^{(l+2)}&= \left(\sum_{i\in \mathcal{S}(\sigma)}(\kappa^{(l+1)}_i)^{\max(1,\beta)}\right)^{1/\max(1,\beta)}\notag\\[3mm]
	&=\left(\sum_{i\in \mathcal{S}(\sigma)}\sum_{s\in \mathcal{S}(i)}(\kappa^{(l)}_s)^{\max(1,\beta)}\right)^{1/\max(1,\beta)}\label{eq:semigroup_kappa}.
\end{alignat}
This final double sum is equivalent to a single sum over all $r^2$ nodes in the unrenormalized layer $l$ that make up the supernode in the layer $l+2$. 

In the similarity dimension we have slightly more freedom, as we just need to find a definition of $\Delta\theta^{(l+1)}_{\sigma\beta}$ that (1) respects the semi-group property of the renormalization procedure,  (2) respects the spherical symmetry of the system and (3) lies in the range defined by the angular coordinates of the constituent nodes and therefore respects the original node order. We therefore define
\begin{equation}
	\theta^{(l+1)}_\sigma=\frac{\sum_{i\in S(\sigma)}(\kappa^{(l)}_i)^{\max(1,\beta)}\theta^{(l)}_i}{\sum_{i\in S(\sigma)}(\kappa^{(l)}_i)^{\max(1,\beta)}}\label{eq:thetaprimeordered},
\end{equation}
which can been seen as a weighted average. Note that we do not choose the exact definition as given in Ref.~\cite{GarciaPerez2018} because it introduces a bias for the constituent node with the largest angular coordinate when $\beta>1$.

The definition in Eq.~\eqref{eq:thetaprimeordered} works well as long as the difference between the largest and smallest coordinate is smaller than $\pi$, but breaks down when this is not the case. If, for example, we try and create a supernode from the coordinates $(\kappa^{(l)}_1,\theta^{(l)}_1)=(1,\pi/4)$ and $(\kappa^{(l)}_2,\theta^{(l)}_2)=(1,7\pi/4)$, we end up with $(\kappa^{(l+1)},\theta^{(l+1)})=(2^{1/\max(1,\beta)},\pi)$. This obviously is not correct, as the supernode lies on the opposite side of the unit circle from where its constituents were located. As we normally define supernodes by taking adjacent constituent nodes, starting from the node with the smallest angular coordinate, we do not run into this problem. However, in the main text we compare the ordered renormalization with one where the constituent nodes are chosen at random, and thus Eq.~\eqref{eq:thetaprimeordered} can in principle not be applied. Note that in this case also the argument used to obtain Eq.~\eqref{eq:renormalized_connection_probability} fails, and so we cannot expect the connection probability in the renormalized layer to be congruent with the $\mathbb{S}^1$ model when the angular coordinate is relevant. The generalization of the renormalized angular coordinate is given by
\begin{equation}
	\theta^{(l+1)}_\sigma=\arg\left(\frac{\sum_{i\in S(\sigma)}(\kappa^{(l)}_i)^{\max(1,\beta)}e^{\mathrm{i}\theta^{(l)}_i}}{\sum_{i\in S(\sigma)}(\kappa^{(l)}_i)^{\max(1,\beta)}}\right)\label{eq:thetaprimerandom},
\end{equation}
which can be seen as a weighted circular mean. 

When the spread of the constituent angular coordinates is small, as is the case for GR, it does not matter which of the two definitions of $\theta^{(l+1)}_\sigma$ one takes: Let $\{\theta^{(l)}_1,...,\theta^{(l)}_r\}$ be the set of constituent nodes of a supernode $\sigma$ in layer $l+1$, sorted in ascending order and where we assume that $\theta^{(l)}_r-\theta^{(l)}_1\ll 1$. Then we know that $\Delta\theta^{(l)}_{i1} \ll 1$ $\forall i$, which allows us to approximate Eq.~\eqref{eq:thetaprimerandom} as 
\begin{alignat}{6}
	\theta^{(l+1)}_\sigma &\approx \arg\left(\frac{e^{i\theta^{(l)}_1}\sum_{i\in\mathcal{S}(\sigma)}\kappa_i^{\max(1,\beta)}\left(1+\mathrm{i}\Delta\theta_{i1}^{(l)}\right)}{\sum_{i\in S(\sigma)}\kappa_i^{\max(1,\beta)}}\right)\notag\\[3mm]
	&=\arg\left(e^{\mathrm{i}\theta^{(l)}_1}\left(1+\mathrm{i}\overline{\Delta\theta^{(l)}}\right)\right)\approx \arg\left(e^{\mathrm{i}\theta_1^{(l)}}e^{\mathrm{i}\overline{\Delta\theta^{(l)}}}\right)\notag\\[3mm]
	&=\theta_1^{(l)} + \overline{\Delta\theta^{(l)}}\label{eq:approximation_theta_l+1},
\end{alignat}
where in the second step we have defined the weighted average of the angular differences
\begin{equation}
	\overline{\Delta\theta^{(l)}}=\frac{\sum_{i\in \mathcal{S}(\sigma)}\kappa_i^{\max(1,\beta)}\left(\theta_i^{(l)}-\theta_1^{(l)}\right)}{\sum_{i\in \mathcal{S}(\sigma)}\kappa_i^{\max(1,\beta)}},
\end{equation}
which is assumed to be small. Eq.~\eqref{eq:approximation_theta_l+1} can be rewritten to obtain Eq.~\eqref{eq:thetaprimeordered}. We generally choose Eq.~\eqref{eq:thetaprimeordered} as it respects the semi-group property explicitly, while Eq.~\eqref{eq:thetaprimerandom} only does so approximately in the case of small angular spread of the constituents as only then can one approximate 
\begin{equation}
	e^{\mathrm{i}\theta_{\sigma}^{(l+1)}}\approx \frac{\sum_{i\in S(\sigma)}(\kappa^{(l)}_i)^{\max(1,\beta)}e^{\mathrm{i}\theta^{(l)}_i}}{\sum_{i\in S(\sigma)}(\kappa^{(l)}_i)^{\max(1,\beta)}},
\end{equation}
which is necessary for the semi-group property to hold.
\\

The expected degree of a node with hidden degree $z=\kappa^{(l+1)}$ can be expressed as
\begin{widetext}
	\begin{alignat}{6}
		\overline{k^{(l+1)}}(z) &= \frac{N^{(l)}}{r}\int\mathrm{d}z'\int_0^\pi\mathrm{d}\theta\rho^{(l+1)}(z')\frac{1}{1+\frac{(R^{(l)}\theta)^\beta}{(\hat{\mu}^{(l)}zz')^{\max(1,\beta)}}}=\frac{N^{(l)}}{r}\int\mathrm{d}z'\rho^{(l+1)}(z'){}_2F_1\left[
		\begin{array}{c}
			1,1/\beta\\
			1+1/\beta\end{array};-\frac{(\pi R^{(l)})^\beta }{(\hat\mu^{(l)}zz')^{\max(1,\beta)}}\right],
	\end{alignat}
\end{widetext}
where we know that for $x\rightarrow\infty$ one has ${}_2F_1(1,1/\beta,1+1/\beta,-x)=((1-\beta)x)^{-1}+(\pi/\beta)\csc(\pi/\beta)x^{-1/\beta}+\mathcal{O}(x^{-2})$. Employing this approximation and using that $\langle \kappa^{(l+1)}\rangle = r^\xi\langle \kappa^{(l)}\rangle$, it can then be shown that 
\begin{equation}
	\overline{k^{(l+1)}}(\kappa^{(l+1)})=r^{\xi-1}\frac{\langle k^{(l)}\rangle}{\langle \kappa^{(l)}\rangle}\kappa^{(l+1)}\label{eq:average_degree_per_degree_class}.
\end{equation}
We can then take the average over the hidden degree to get
\begin{equation}
	\langle k^{(l+1)}\rangle = r^{\nu}\langle k^{(l)}\rangle\label{eq:evolution_average_k},
\end{equation}
where we define $\nu=2\xi-1$.

\section{Self-similarity of the degree distribution}
The goal of this section is to find the degree distribution at the $r$'th level of renormalization. We start by studying the hidden degree distribution, assuming that in the original network the distribution is given by
\begin{equation}
	\rho(\kappa) = \mathcal{N}\kappa^{-\gamma},\quad\quad\kappa_0\leq\kappa\leq\kappa_c\label{eq:appendix_kappa_distribution},
\end{equation}
where $\mathcal{N}$ is the normalization constant. To obtain the distribution after renormalization we use Eq.~\eqref{eq:kappaevol}. Note that we change our method slightly from this point onward. Instead of looking at the $l$'th layer of the iterative normalization procedure where each supernode is constructed with $r$ nodes, we now study only a single normalization step. Note however that, due to the semi-group property, these two approaches are equivalent as $l$ steps of size $r$ can always be replaced by a single step of size $r^l$. We first find that the distribution $\tilde\rho(\tilde\kappa)$, where $\tilde\kappa=\kappa^{\max(1,\beta)}$:
\begin{equation}
	\tilde\rho(\tilde\kappa) = \tilde{\mathcal{N}} \tilde\kappa^{-\eta},\quad\quad \tilde\kappa_0\leq \tilde\kappa \leq  \tilde\kappa_c,
\end{equation}
where we have defined $\tilde{\mathcal{N}}=\mathcal{N}/\max(1,\beta)$, $\tilde\kappa_0=\kappa_0^{\max(1,\beta)}$, $\tilde\kappa_c=\kappa_c^{\max(1,\beta)}$ and $\eta=1+(\gamma-1)/\max(1,\beta)$. The next step is to find the distribution $\tilde\rho_r(\tilde z)$ where $\tilde z=\sum_{i=1}^r\tilde\kappa_i$. We first state the result and follow with the proof:
\begin{equation}
	\tilde\rho_{r}(\tilde z) = \sum_{n=1}^r\sum_{q=1}^{\infty}c_{n,q}\tilde z^{n(1-\eta)-q}1_{[r\tilde\kappa_0,\tilde\kappa_c+(r-1)\tilde\kappa_0]}(\tilde z)\label{eq:tilderhor},
\end{equation}
where $c_{n,l}$ are constants. To obtain the distribution of the hidden degrees $z$ in the renormalized layer, we use the fact that $z=\tilde z^{1/\max(1,\beta)}$, which leads to 
\begin{equation}
	\rho_r(z)=\max(1,\beta)\tilde\rho_r(z^{\max(1,\beta)})z^{\max(1,\beta)-1}\label{eq:finaltransform}.
\end{equation}
Note that for $\tilde z\gg 1$, the dominant scaling in Eq.\eqref{eq:tilderhor} is $\tilde z^{-\eta}$ ($n=1,q=1$). Plugging this into Eq.\eqref{eq:finaltransform} proves that the distribution $\rho_r(z)$ scales as $z^{-\gamma}$, which in turn demonstrates the self-similarity of the scaling behavior of the hidden degree distribution under renormalization. Note that the cut-off in the renormalized layer is given by $(\tilde\kappa_c+(r-1)\tilde\kappa_0)^{1/\max(1,\beta)}$, which is approximately $\kappa_c$ if $\tilde\kappa_c\gg(r-1)\tilde\kappa_0$.

We now prove Eq.~\eqref{eq:tilderhor} using induction. First, for $r=2$, we know that the distribution $\tilde\rho_2(\tilde z)$, where $\tilde z=\tilde \kappa_1+\tilde \kappa_2$, is given by the convolution
\begin{equation}
	\tilde\rho_2(\tilde z) = \int_{-\infty}^\infty \mathrm{d}\tilde \kappa\tilde \rho(\tilde z-\tilde \kappa)\tilde \rho(\tilde \kappa)\label{eq:convolution}.
\end{equation}
Taking into account the support of $\tilde\rho(\tilde \kappa)$ we can conclude that $\tilde \kappa_0\leq \tilde z-\tilde \kappa\leq\tilde \kappa_c$ and $\tilde \kappa_0\leq\tilde \kappa\leq\tilde \kappa_c$.  We then rewrite Eq.\@~\ref{eq:convolution} as
\begin{widetext}
	\begin{alignat}{6}
		\tilde \rho_2(\tilde z) &= \frac{\tilde {\mathcal{N}}^2}{\tilde z^{2\gamma-1}}\Bigg[\bigg(B_{1-\frac{\tilde \kappa_0}{\tilde z}}
		\begin{bmatrix}
			1-\eta\\
			1-\eta
		\end{bmatrix}	
		-B_{\frac{\tilde \kappa_0}{\tilde z}}
		\begin{bmatrix}
			1-\eta\\
			1-\eta
		\end{bmatrix}\bigg)
		1_{[2\tilde \kappa_0,\tilde \kappa_c+\tilde \kappa_0]}(\tilde z)
		+
		\bigg(B_{\frac{\tilde \kappa_c}{\tilde z}}
		\begin{bmatrix}
			1-\eta\\
			1-\eta
		\end{bmatrix}
		-
		B_{1-\frac{\tilde \kappa_c}{\tilde z}}
		\begin{bmatrix}
			1-\eta\\
			1-\eta
		\end{bmatrix}
		\bigg)1_{[\tilde \kappa_0+\tilde \kappa_c,2\tilde \kappa_c]}(\tilde z)\Bigg].
	\end{alignat}
\end{widetext}
Here, $B_a\begin{bmatrix}b\\c\end{bmatrix}$ represents the incomplete beta function. We then note that this function can be expanded as
\begin{alignat}{6}
	B_{1-x}\begin{bmatrix}a\\b\end{bmatrix}&=\frac{\pi\csc{(b\pi)}}{a}\left(\sum_{n=0}^\infty\frac{a_{(n)}}{n!}(-x)^n\right)
	\times\bigg(\frac{\Gamma(1+a)}{\Gamma(a+b)}\sum_{q=0}^\infty\left[\frac{(b-1)_{(q)}(-a)_{(q)}}{q!\Gamma(1-b+q)}x^l\right]\notag\\[3mm]
	&-\frac{ax^b}{\Gamma(1-b)}\sum_{q=0}^\infty\left[\frac{(-a-b)_{(q)}}{q!\Gamma(1+b+q)}(-x)^q\right]\bigg)\\
	B_x\begin{bmatrix}a\\b\end{bmatrix}&=x^a\sum_{n=0}^\infty\frac{(1-b)_{(n)}}{n!(a+n)}x^n
\end{alignat}
when $x\rightarrow0$, where the $y_{(n)}$ represent the falling factorials: $y_{(n)}=y(y-1)(y-2)...(y-n+1)$. In the case that $\tilde z\in[2\tilde \kappa_0,\tilde \kappa_c+\tilde \kappa_0]$, $\tilde z/\tilde \kappa_0\ll1$ in the tail of the distribution. Thus, we can apply the expansions given above and show that the dominant scaling in this regime is $\tilde \rho_2(\tilde z)\sim \tilde z^{-\eta}$ and that the full behavior is given by Eq.~\eqref{eq:tilderhor}. Crossing over to the regime $\tilde z\in[\tilde \kappa_0+\tilde \kappa_c,2\tilde \kappa_c]$, we get that $1-\tilde \kappa_c/\tilde z\ll1$, as least close to the transition. Using once again the series expansions of the beta functions we obtain that $\tilde \rho_2(z)\sim(1-\tilde \kappa_c/\tilde z)^{1-\eta}$. This falls of hyperbolically and so we can take the probability density to be zero here. Therefore, we prove Eq.~\eqref{eq:tilderhor} for $r=2$. 

Now, assuming that Eq.~\eqref{eq:tilderhor} is true for some general $r$, let us investigate the case for $r+1$. In this case, we start with the convolution 
\begin{alignat}{6}
	&\tilde \rho_{r+1}(\tilde z)=\int_{r\tilde \kappa_0}^{\tilde z-\tilde \kappa_0}\mathrm{d}\tilde \kappa\rho_1(\tilde z-\tilde \kappa)\rho_r(\tilde \kappa)1_{[(r+1)\tilde \kappa_0,\tilde \kappa_c+r\tilde \kappa_0]}(\tilde z)\notag\\[3mm]
	&+\int_{\tilde z-\tilde \kappa_c}^{\tilde \kappa_c+(r-1)\tilde \kappa_0}\mathrm{d}\tilde \kappa\tilde \rho_1(\tilde z-\tilde \kappa)\tilde \rho_r(\tilde \kappa)1_{[\tilde \kappa_c+r\tilde \kappa_0,2\tilde \kappa_c+(r-1)\tilde \kappa_0]}(\tilde z),
\end{alignat}
where have taken into account the respective domains of the two functions $\tilde \rho_1$ and $\tilde \rho_r$. The fact that $\tilde \rho_r(\tilde z)$ can be expanded into a sum of terms $\tilde g_r(\tilde z;\alpha)\sim \tilde z^{-\alpha}$, where $\alpha\geq\eta$, implies that $\tilde \rho_{r+1}(\tilde z)$ can be expanded into a sum of integrals $\tilde I(\tilde z;\alpha)$ evaluating to 
\begin{widetext}
	\begin{alignat}{6}
		\tilde I(\tilde z;\alpha)&&=\tilde {\mathcal{N}}^r\,\tilde z^{1-\eta-\alpha}\,\bigg[\bigg(B_{1-\frac{\tilde \kappa_0}{\tilde z}}\begin{bmatrix}1-\alpha\\1-\eta\end{bmatrix}-&&B_{\frac{r\tilde \kappa_0}{\tilde z}}\begin{bmatrix}1-\alpha\\1-\eta\end{bmatrix}&\bigg)1_{[(r+1)\tilde \kappa_0,\tilde \kappa_c+r\tilde \kappa_0]}\notag\\[3mm]
		&&+\,\,\bigg(B_{\frac{\tilde \kappa_c+(r-1)\tilde \kappa_0}{\tilde z}}\begin{bmatrix}1-\alpha\\1-\eta\end{bmatrix}-&&B_{1-\frac{\tilde \kappa_c}{\tilde z}}\begin{bmatrix}1-\alpha\\1-\eta\end{bmatrix}&\bigg)1_{[\tilde \kappa_c+r\tilde \kappa_0,2\tilde \kappa_c+(r-1)\tilde \kappa_0]}\bigg].
	\end{alignat}
\end{widetext}
Using the same arguments as before, we can show that $\forall\alpha$ the integral falls off hyperbolically in the second region. When $\tilde z\in[(r+1)\tilde \kappa_0,\tilde \kappa_c+r\tilde \kappa_0]$, it can be shown that the expression can be rewritten in the form of Eq.~\eqref{eq:tilderhor}, where the dominant scaling for large $\tilde z$ is once again $\sim \tilde z^{-\eta}$. With this we conclude the proof.

Note that this proof is contingent on some assumptions, most notably that $r\tilde \kappa_0\ll\tilde \kappa_c$. Of course, for finite $\tilde \kappa_c$, there is always an $r$ for which this assumption breaks down. This has to do with the central limit theorem: For a finite cut-off $\tilde \kappa_c$, the variance of the distribution $\tilde \rho(\tilde \kappa)$ is also finite, and thus the distribution $\tilde \rho_r(\tilde z)$ necessarily approaches a Gaussian as $r\rightarrow\infty$. In the case of the model we in general assume that $\tilde\kappa_c=\tilde\kappa_0N^{\max(1,\beta)/(\gamma-1)}$, which is very large for the network sizes we typically work with, and so one can perform several renormalization steps before one `feels' the effect of the cut-off. \\

It is known that the degree distribution is related to the distribution of hidden degrees by 
\begin{equation}
	P_r(k)=\frac{1}{k!}\int\mathrm{d}z\rho(z)\overline{k}(z)^ke^{-\overline k(z)}\label{eq:Pkgeneral},
\end{equation}
where $\overline{k}(\kappa)$ is the expected degree of a node with hidden degree $\kappa$. In the unrenormalized layer one can show that $\overline{k}(\kappa)=\kappa$ when $\hat\mu$ is chosen correctly. For this to be true for in the renormalized layer, however, one would need that $\langle \kappa_r\rangle = \langle k_r\rangle$, which is not generally the case as the scaling exponents determining the flow of these two quantities, $\xi$ and $\nu$, are not always equal. Using  Eq.~\eqref{eq:average_degree_per_degree_class} and $\xi=(\nu+1)/2$, one obtains that 
\begin{equation}
	\overline{k_r}(\kappa_r) = r^{(\nu-1)/2}\kappa_r
\end{equation}
We now note that we do not know the exact functional form of $\rho_r(\kappa)$, at least not for $\beta>1$. To be able to plug in Eq.~\eqref{eq:tilderhor}, we first need to transform \eqref{eq:Pkgeneral}. It can be shown that this integral is equivalent to
\begin{equation}
	P_r(k)=\frac{1}{k!}\int_{r\tilde\kappa_0}^{\tilde\kappa_c+(r-1)\tilde\kappa_0}\mathrm{d}\tilde z\frac{r^{k(\nu-1)/2}\tilde\rho_r(\tilde z)\tilde z^{\frac{k}{\max(1,\beta)}}}{\exp\big(r^{(\nu-1)/2}\tilde z^{\frac{1}{\max(1,\beta)}}\big)}.
\end{equation}
Then, combining the previous result with Eq.~\eqref{eq:tilderhor} and Eq.~\eqref{eq:Pkgeneral} one obtains 
\begin{widetext}
	\begin{alignat}{6}
		P_r(k)&=\sum_{n=1}^r\sum_{q=1}^{\infty}\frac{c_{n,q}r^{k\frac{\nu-1}{2}}}{k!}\int_{r\tilde\kappa_0}^{\tilde \kappa_c+(r-1)\tilde\kappa_0}\mathrm{d}\tilde z \frac{\tilde z^{n(1-\eta)-q+k/\max(1,\beta)}}{\exp\left(r^{\frac{\nu-1}{2}}\tilde z^{1/\max(1,\beta)}\right)}\notag\\[3mm]
		&=\sum_{n=1}^r\sum_{q=1}^{\infty}\frac{c_{n,l}\max(1,\beta)r^{\frac{\nu-1}{2}(\max(1,\beta)(1-q)+n(1-\gamma))}}{k!}\Bigg[\Gamma\left(\max(1,\beta)(1-q)+n(1-\gamma)+k,r^{\frac{\nu-1}{2}+1/\max(1,\beta)}\kappa_0\right)\notag\\[3mm]
		&+\Gamma\left(\max(1,\beta)(1-q)+n(1-\gamma)+k,r^{\frac{\nu-1}{2}}\left(\tilde\kappa_c+(r-1)\tilde\kappa_0\right)^{1/\max(1,\beta)}\right)\Bigg].
	\end{alignat}
\end{widetext}
When $k\gg\left(\tilde\kappa_c+(r-1)\tilde\kappa_0\right)^{1/\max(1,\beta)}$, the two gamma functions cancel, meaning that the probability density vanishes.  When $r^{1/\max(1,\beta)}\kappa_0\ll k\leq\left(\tilde\kappa_c+(r-1)\tilde\kappa_0\right)^{1/\max(1,\beta)}$, the first term scales as $k^{-\gamma}$, whereas the second term falls off exponentially. This implies that the scaling behavior of the tail of the distribution is preserved under renormalization. Note that once again for large $\kappa_c$ the  cut-off does not evolve under renormalization.

\section{Real Networks}
In this section we show the results of the RG procedure applied to a set of Real Networks living in the weakly geometric region $\beta<1$. We present here short descriptions of these networks as given in Ref.~\cite{vanderkolk2024}. The properties of these networks are shown in Tab.~\ref{tab:realnets}. 

\begin{itemize}
	\item \textbf{Foodweb--Eocene}~\cite{Dunne2014}: A reconstructed food web of an ecosystem from the early Eocene (48 million years ago). Nodes represent taxa and edges represent consumer-resource relations. The original network was directed.  
	\item \textbf{WordAdjacency--English}~\cite{Milo2004}: A network of word adjacency in English texts. Nodes represent words and two words are connected if one directly follows the other in texts. The original network was directed. 
	\item \textbf{WordAdjacency--Japanese}~\cite{Milo2004}: A network of word adjacency in Japanese texts. Nodes represent words and two words are connected if one directly follows the other in texts. The original network was directed.
	\item \textbf{MB--R.norvegicus}~\cite{Huss2007}: A metabolic network of the rat (Ratus norvegicus), extracted from the Kyoto Encyclopedia of Genes and Genomes (KEGG). Nodes represent substances involved in enzymatic reactions and edges represent reactant-product pairs. 
	\item \textbf{WikiTalk--Catalan}~\cite{Kunegis2013}: A network where nodes represents Wikipedia editors for a certain language (in this case Catalan), and where user $i$ and $j$ are connected if $i$ leaves a message on the talk page of $j$. The original network was directed. 
	\item \textbf{GI--S.cerevisiae}~\cite{Hu2018}: A network based on the Molecular Interaction Search Tool (MIST) for baker's yeast (Saccharomyces cerevisiae). Here node represent genes and the edges indicate that the effects of mutations in one gene can be modified by mutations of another gene. 
	\item \textbf{GMP--C.elegans}~\cite{DeDomenico2015}: A multiplex network representing different types of genetic interactions for the nematode worm Caenorhabditis elegans. The layers represent physical, association, co-localization, direct, suppressive and additive interactions. In this paper we create a monolayer network by treating the different interaction types equally and removing repeated links. The original network was directed. 
	\item \textbf{Gnutella}~\cite{Ripeanu2002}: A snapshot of the Gnutella peer-to-peer file sharing network on August 4th 2002. Nodes are hosts and edges are connections between them. The original network was directed. 
	\item  \textbf{PPI--S.cerevisiae}~\cite{Hu2018}: A network based on the Molecular Interaction Search Tool (MIST) for baker's yeast (Saccharomyces cerevisiae). Here node represent genes and the edges indicate that there are physical interactions between their associated proteins. 
	\item \textbf{PPI--D.melanogaster}~\cite{Hu2018}: A network based on the Molecular Interaction Search Tool (MIST) for the fruit fly (Drosophila melanogaster). Here node represent genes and the edges indicate that there are physical interactions between their associated proteins.
	\item \textbf{Transport--London}~\cite{DeDomenico2014}: An multiplex network of the public transportation system in London. Nodes are London train stations and the links can represent either the underground, overground and DLR connections. There connections are treated equally as to create a mono-layer network. 
	\item \textbf{GMP--S.cerevisiae}~\cite{DeDomenico2015}: A multiplex network representing different types of genetic interactions for baker's yeast (Saccharomyces cerevisiae). The layers represent physical, association, co-localization, direct, suppressive and additive interactions. In this paper we create a monolayer network by treating the different interaction types equally and removing repeated links. The original network was directed. 
	\item \textbf{Internet-PoP}~\cite{Knight2011}: The Kentucky Datalink network, an internet graph at the Point of Presence (PoP) level. Nodes are physical network interface points and links physical connections between them.
	\item \textbf{PPI--H.sapiens}~\cite{Hu2018}: A network based on the Molecular Interaction Search Tool (MIST) for humans (Homo sapiens). Here node represent genes and the edges indicate that there are physical interactions between their associated proteins.
	\item \textbf{WikiVote}~\cite{Leskovec2010}: The network represents the voting process used to select Wikipedia administrators, which are contributors with access to additional technical features. Nodes represents Wikipedia users and an edge is created if user $i$ votes on the selections of user $j$. The original network was directed.
	\item \textbf{MathOverflow}~\cite{Paranjape2017}: An interaction network of users (nodes) on the online Q\&A site MathOverflow. An edge from node $i$ to node $j$ indicates that $i$ responded to an answer by $j$. The original network was directed. 
\end{itemize}
\begin{table*}[h]
	\caption{\label{tab:realnets} Network properties of several real weakly geometric networks shown. The following abbreviations are used: (MB) Metabolic, (GI) Genetic Interactions, (GMP) Genetic Multiplex, (PPI) Protein Protein Interactions, (PoP) Point of Presence.  }
	\begin{ruledtabular}
		\begin{tabular}{lcccccc}
			\textbf{Network} 			& \textbf{Area}& \textbf{$N$} & \textbf{$\langle k\rangle$} & $k_{\text{max}}$ &  \textbf{$\overline{c}$} & \textbf{$\beta$}  \\\hline
			Foodweb--Eocene 				&Ecological		& $700$		&$18.3$		&$192$	&$0.10$&$\beta\approx0$\\
			WordAdjacency--English 	    & Language 		&$7377$		&$12.0$		&$2568$	&$0.47$&$\beta\approx0$\\
			WordAdjacency--Japanese	    & Language		& $2698$ 	& $5.9$ 	&$725 $ 	& $0.30$ & $\beta\approx0$\\
			MB--R.norvegicus			    &Cell 			& $1590$ 	& $5.9$ 	&$498 $ 	& $0.19$ & $\beta\approx0$\\ 
			WikiTalk--Catalan 			&Social			& $79209$ 	& $4.6$ 	& $53234 $ 	& $0.83$ & $\beta\approx0$\\ 
			GI--S.cerevisiae			    &Cell			&$5933$ 	& $149$ 	& $2244$ & $0.17$ & $0.63$\\ 
			GMP--C.elegans 				&Cell		  	& $3692$	&$4.1$		&$526$	&$0.11$&$0.69$\\
			Gnutella 					&Technological	& $10876$	& $7.4$ 	& $103$ & $0.01$ & $0.73$ \\
			PPI--S.cerevisiae 			&Cell			& $7271$ 	& $45.0$ 	& $3613$ & $0.37$ & $0.75$ \\ 
			PPI--D.melanogaster 			&Cell 			& $11319$ 	& $23.7$ 	& $889 $ 	& $0.10$ & $0.84$ \\
			Transport--London		    &Transportation	& $369$ 	& $2.3$ 	&$7$	& $0.03$ & $0.86$ \\ 
			GMP--S.cerevisiae		    &Cell			& $6567$	&$68.1$		&$3254$	&$0.22$&$0.88$\\
			Internet-PoP 				&Technological	& $754$ 	& $2.4$ 	& $7$ & $0.03$ & $0.90$ \\
			PPI--H.sapiens				&Cell			&$27578$	&$37.9$		&$2883$ &$0.15$	&$0.91$\\
			WikiVote				    &Social			& $7066$ 	& $28.5$ 	& $1065$ & $0.21$ & $0.91$\\
			MathOverflow 				&Social			& $13599$ 	& $10.5$ 	& $949 $ 	& $0.32$ & $0.99$ \\
		\end{tabular}
	\end{ruledtabular}
\end{table*}
\begin{figure}[h]
	\centering
	\includegraphics[width=1\textwidth]{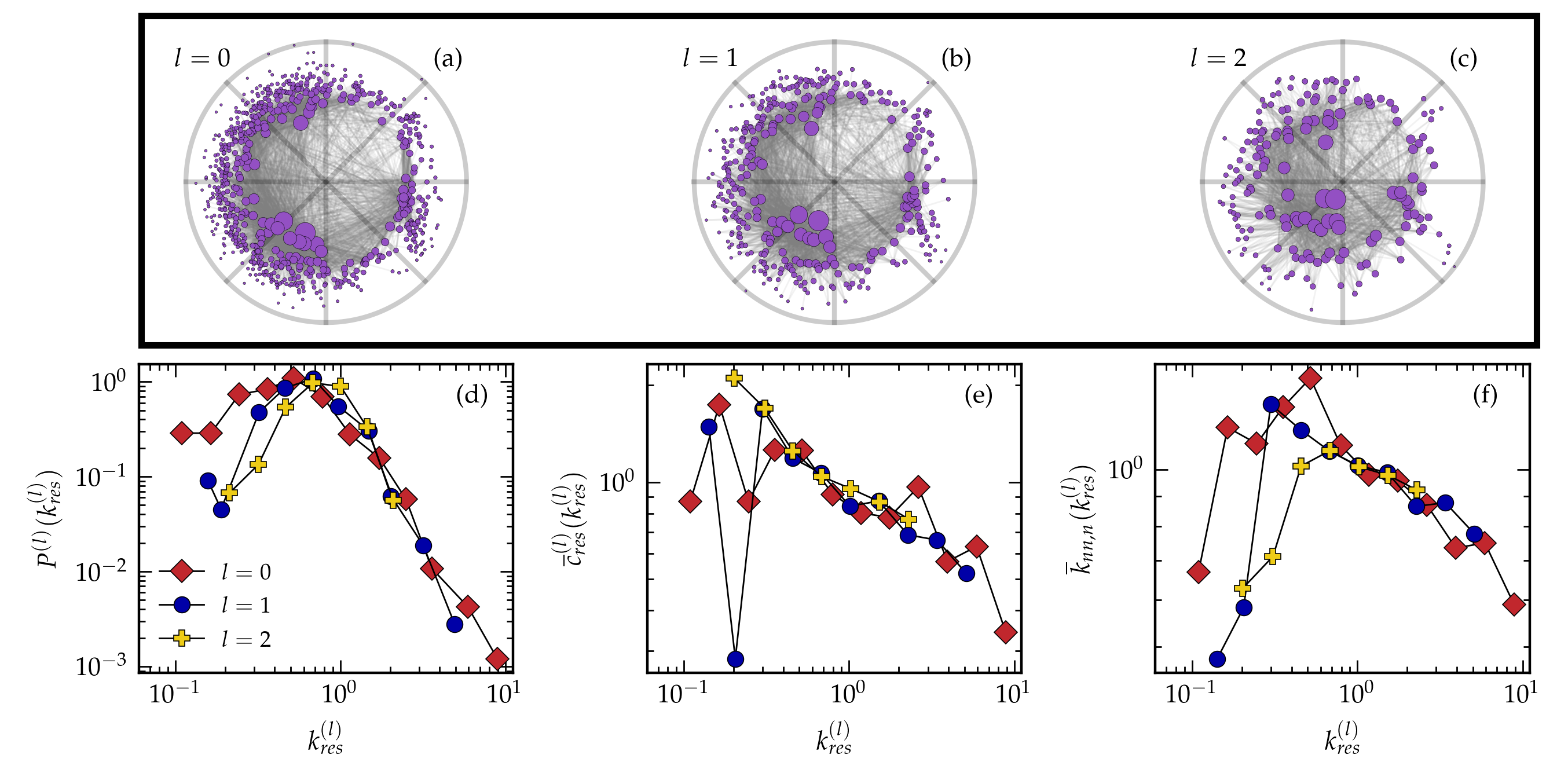}
	\vspace{-8mm}
	\caption{Summary of the results of GR for the Foodweb-Eocene network. \textbf{(a-c)} Representation of the embedding for layers $l=1,2 $ and $3$ in the hyperbolic plane. The top 50\% most geometric edges are shown. The topological properties are also given: \textbf{(d)} the degree distribution, where $k_{res}^{(l)}=k^{(l)}/\langle k^{(l)}\rangle$, \textbf{(e)} the rescaled average local clustering coefficient per degree class, where $\overline c^{(l)}_{res}(k^{(l)}) = \overline c^{(l)}(k^{(l)})/\overline c^{(l)}$ and finally \textbf{(f)} the degree-degree correlations per degree class. In all cases we log-bin the degrees.  }
	\label{Sfig:foodweb_eocene_R}
\end{figure}
\begin{figure}[b]
	\centering
	\includegraphics[width=1\textwidth]{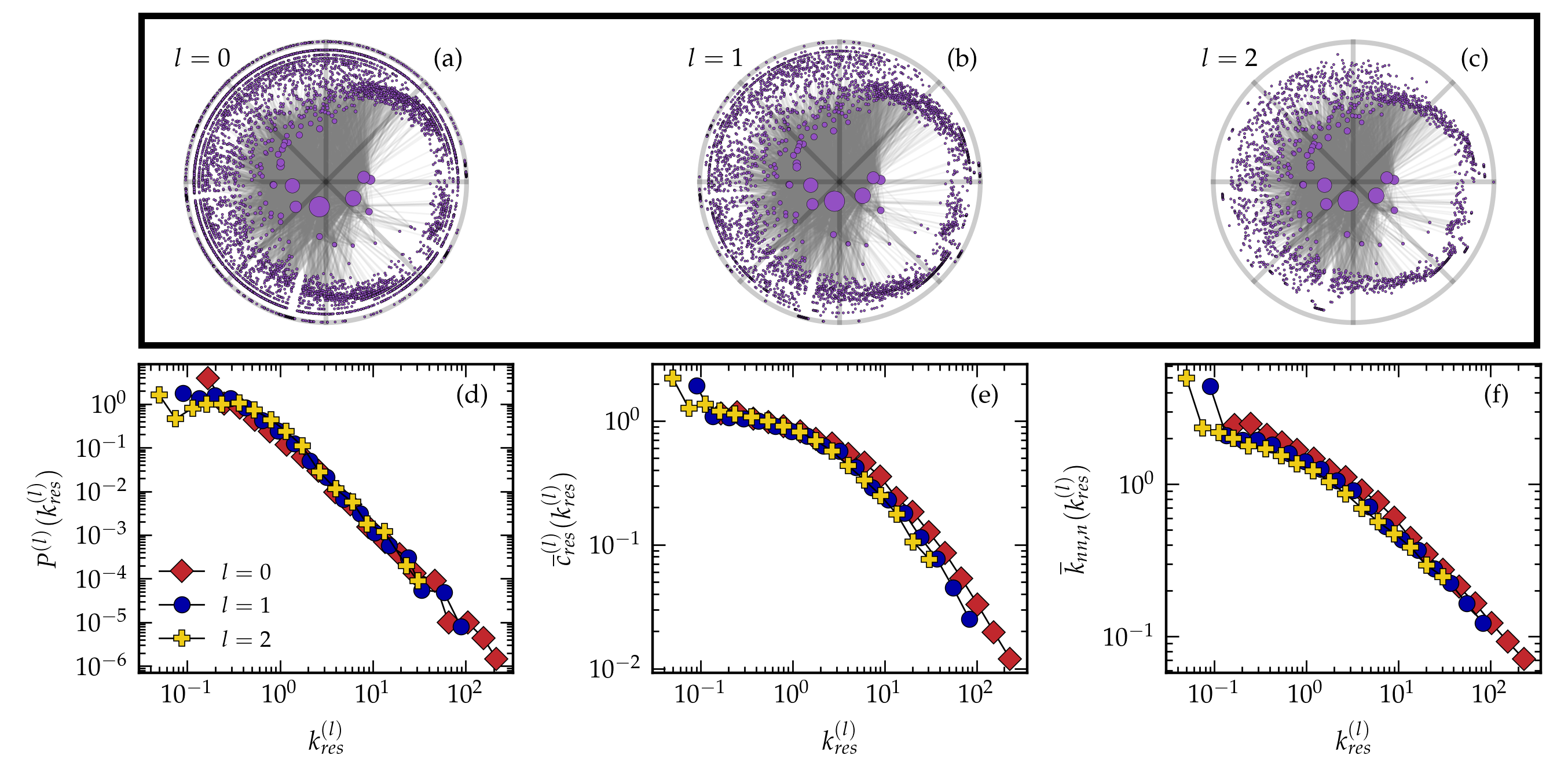}
	\vspace{-8mm}
	\caption{Summary of the results of GR for the WordAdjacency-English network. \textbf{(a-c)} Representation of the embedding for layers $l=1,2 $ and $3$ in the hyperbolic plane. The top 10\% most geometric edges are shown. The topological properties are also given: \textbf{(d)} the degree distribution, where $k_{res}^{(l)}=k^{(l)}/\langle k^{(l)}\rangle$, \textbf{(e)} the rescaled average local clustering coefficient per degree class, where $\overline c^{(l)}_{res}(k^{(l)}) = \overline c^{(l)}(k^{(l)})/\overline c^{(l)}$ and finally \textbf{(f)} the degree-degree correlations per degree class. In all cases we log-bin the degrees.  }
	\label{Sfig:WordAdjacency_English_R}
\end{figure}
\begin{figure}[h]
	\centering
	\includegraphics[width=1\textwidth]{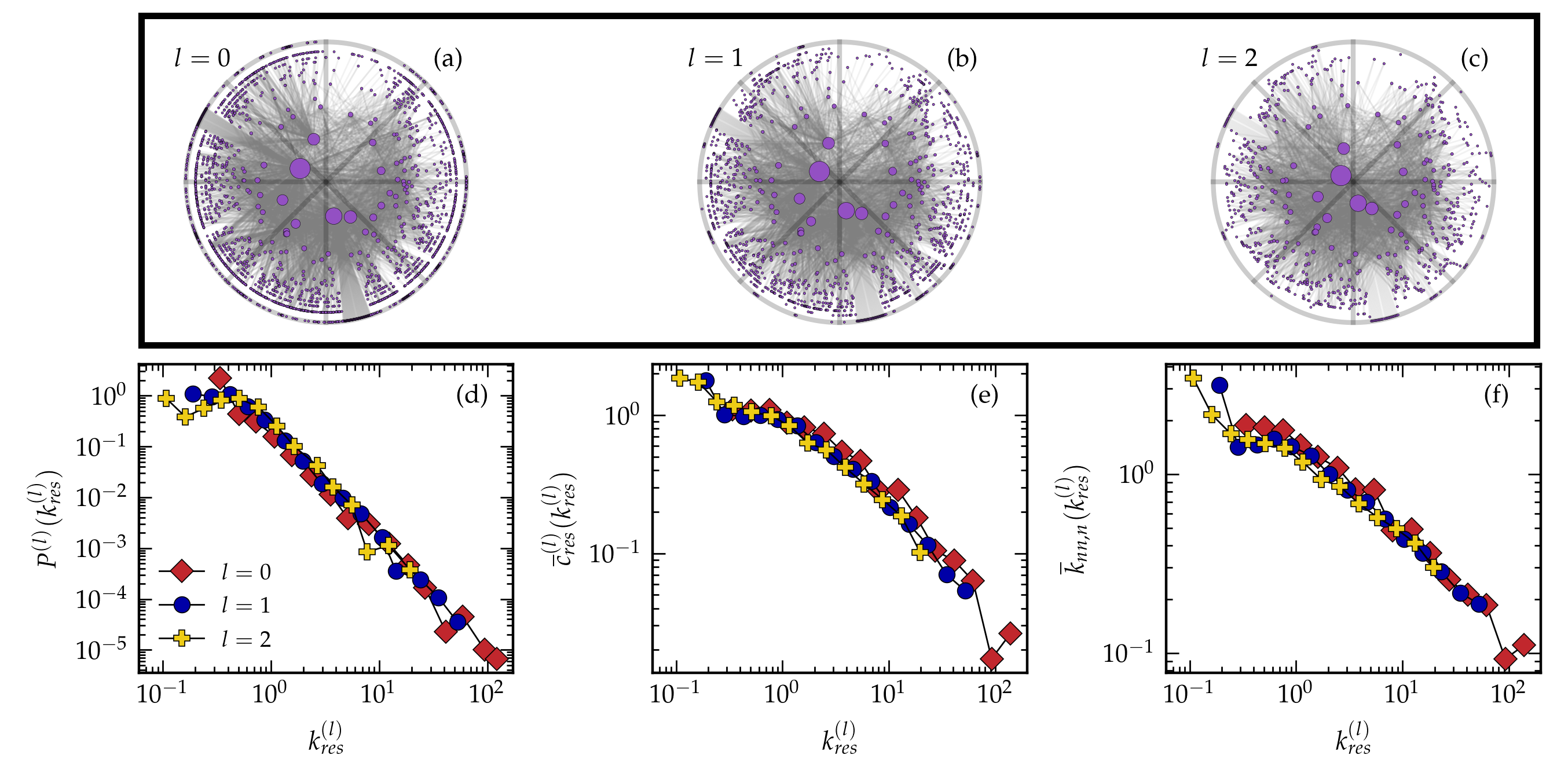}
	\vspace{-8mm}
	\caption{Summary of the results of GR for the WordAdjacency–Japanese network. \textbf{(a-c)} Representation of the embedding for layers $l=1,2 $ and $3$ in the hyperbolic plane. The top 50\% most geometric edges are shown. The topological properties are also given: \textbf{(d)} the degree distribution, where $k_{res}^{(l)}=k^{(l)}/\langle k^{(l)}\rangle$, \textbf{(e)} the rescaled average local clustering coefficient per degree class, where $\overline c^{(l)}_{res}(k^{(l)}) = \overline c^{(l)}(k^{(l)})/\overline c^{(l)}$ and finally \textbf{(f)} the degree-degree correlations per degree class. In all cases we log-bin the degrees.  }
	\label{Sfig:WordAdjacency_Japanese_R}
\end{figure}
\begin{figure}[h]
	\centering
	\includegraphics[width=1\textwidth]{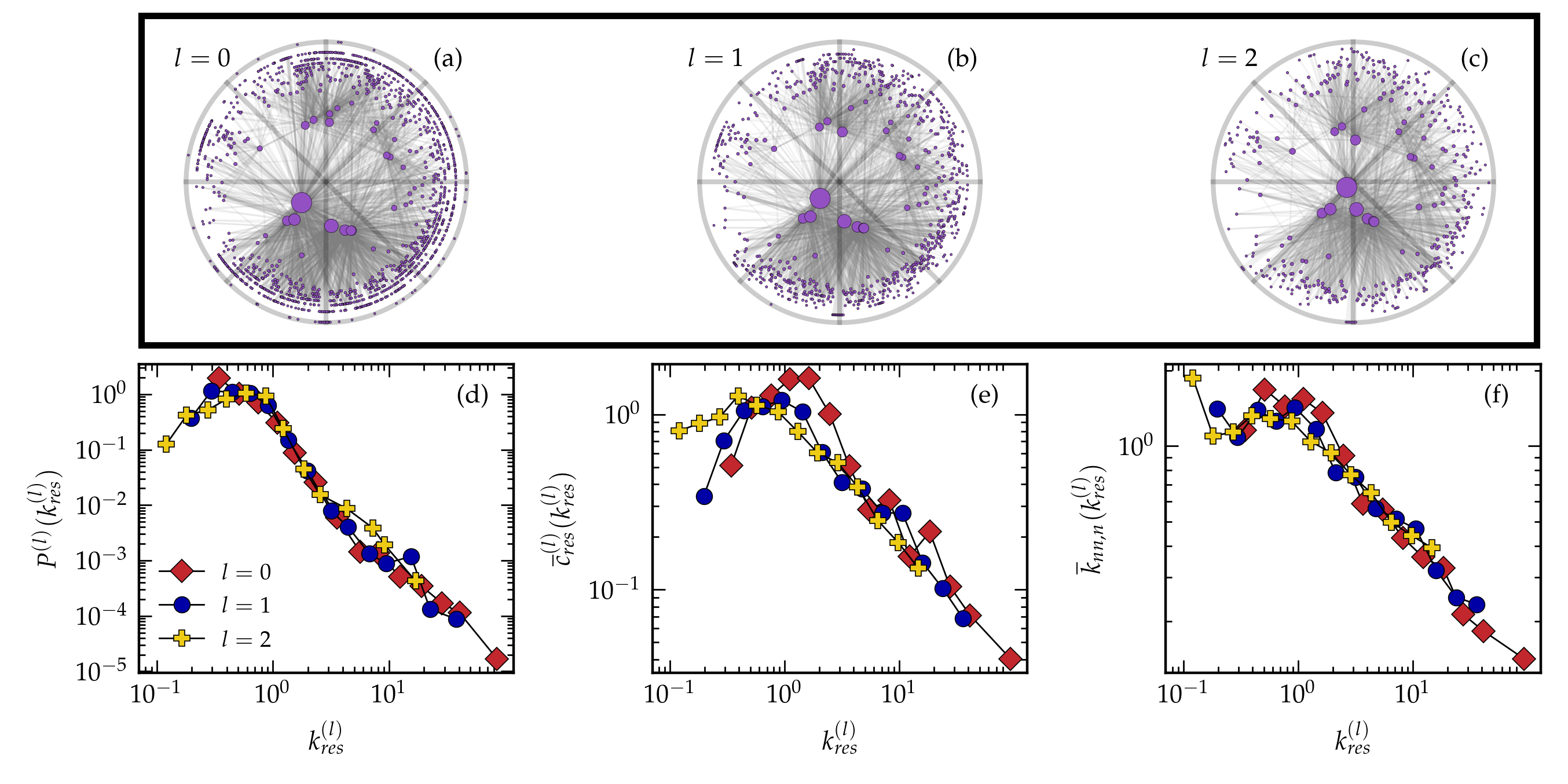}
	\vspace{-8mm}
	\caption{Summary of the results of GR for the MB–R.norvegicus network. \textbf{(a-c)} Representation of the embedding for layers $l=1,2 $ and $3$ in the hyperbolic plane. The top 50\% most geometric edges are shown. The topological properties are also given: \textbf{(d)} the degree distribution, where $k_{res}^{(l)}=k^{(l)}/\langle k^{(l)}\rangle$, \textbf{(e)} the rescaled average local clustering coefficient per degree class, where $\overline c^{(l)}_{res}(k^{(l)}) = \overline c^{(l)}(k^{(l)})/\overline c^{(l)}$ and finally \textbf{(f)} the degree-degree correlations per degree class. In all cases we log-bin the degrees.  }
	\label{Sfig:MB-R.norvegicus}
\end{figure}
\begin{figure}[h]
	\centering
	\includegraphics[width=1\textwidth]{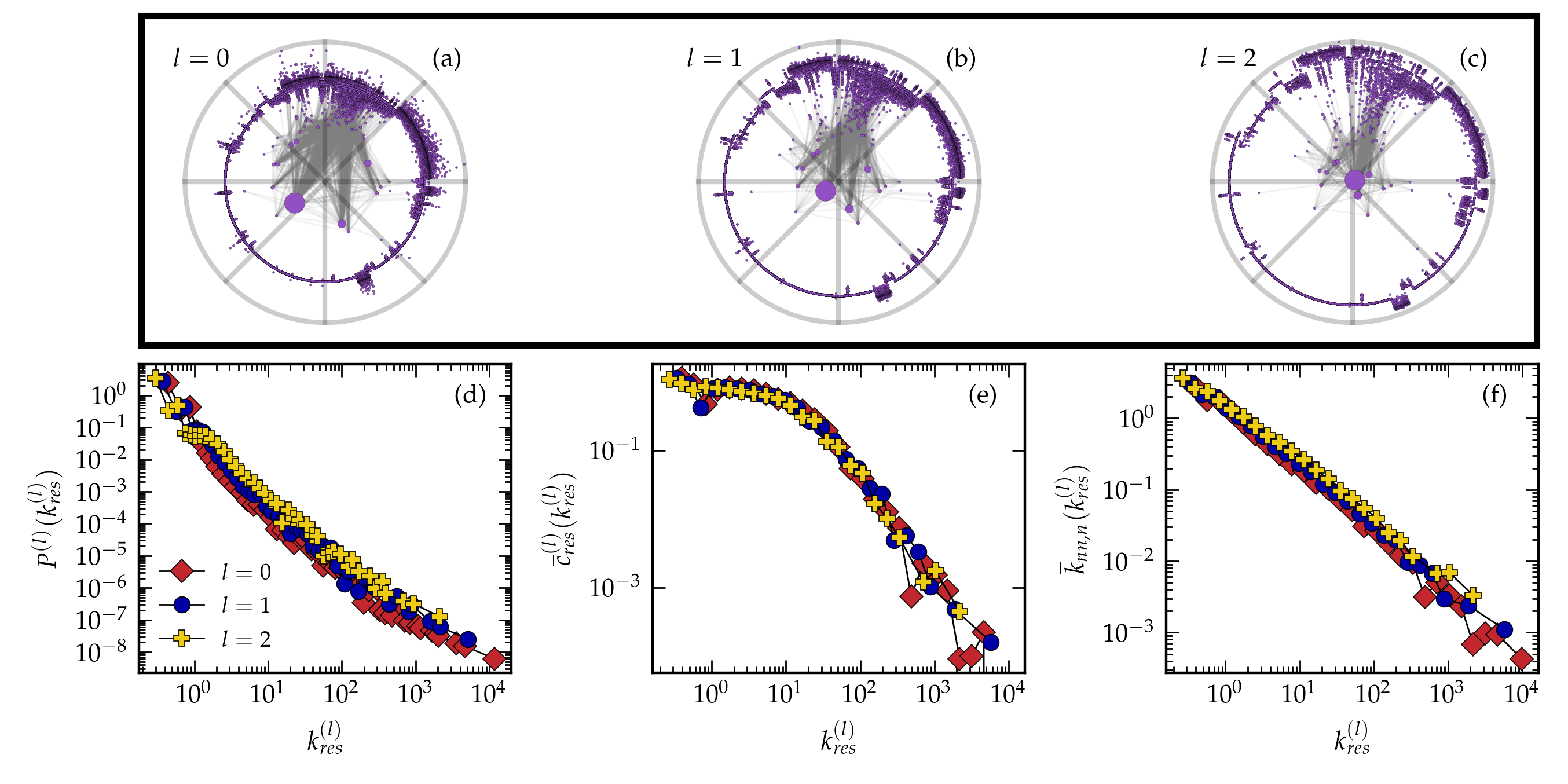}
	\vspace{-8mm}
	\caption{Summary of the results of GR for the WikiTalk–Catalan network. \textbf{(a-c)} Representation of the embedding for layers $l=1,2 $ and $3$ in the hyperbolic plane. The top 1\% most geometric edges are shown. The topological properties are also given: \textbf{(d)} the degree distribution, where $k_{res}^{(l)}=k^{(l)}/\langle k^{(l)}\rangle$, \textbf{(e)} the rescaled average local clustering coefficient per degree class, where $\overline c^{(l)}_{res}(k^{(l)}) = \overline c^{(l)}(k^{(l)})/\overline c^{(l)}$ and finally \textbf{(f)} the degree-degree correlations per degree class. In all cases we log-bin the degrees.  }
	\label{Sfig:WikiTalk_Catalan_R}
\end{figure}
\begin{figure}[h]
	\centering
	\includegraphics[width=1\textwidth]{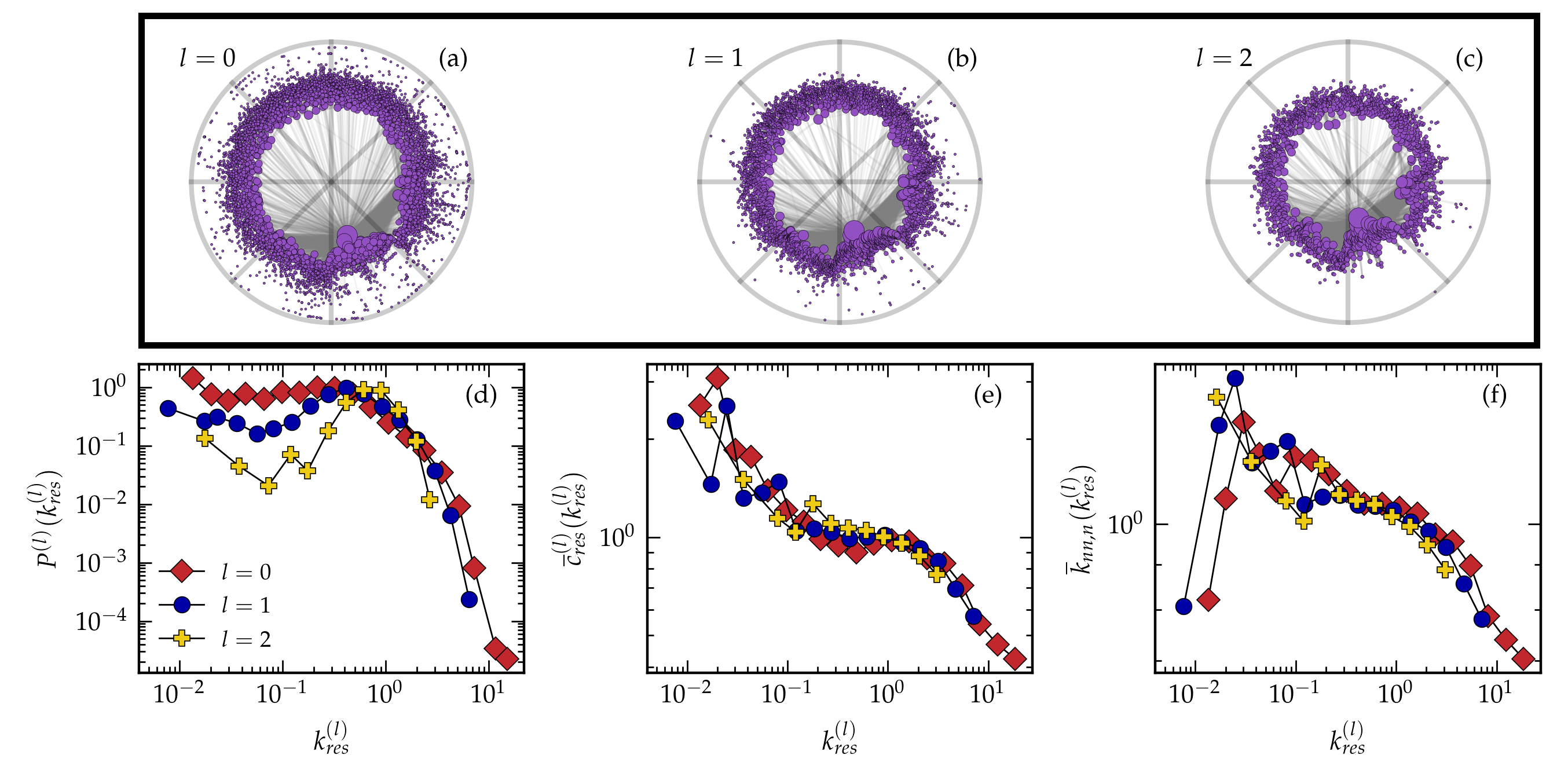}
	\vspace{-8mm}
	\caption{Summary of the results of GR for the GI–S.cerevisiae network. \textbf{(a-c)} Representation of the embedding for layers $l=1,2 $ and $3$ in the hyperbolic plane. The top 5\% most geometric edges are shown. The topological properties are also given: \textbf{(d)} the degree distribution, where $k_{res}^{(l)}=k^{(l)}/\langle k^{(l)}\rangle$, \textbf{(e)} the rescaled average local clustering coefficient per degree class, where $\overline c^{(l)}_{res}(k^{(l)}) = \overline c^{(l)}(k^{(l)})/\overline c^{(l)}$ and finally \textbf{(f)} the degree-degree correlations per degree class. In all cases we log-bin the degrees.  }
	\label{Sfig:GI-Scerevisiae_R}
\end{figure}
\begin{figure}[h]
	\centering
	\includegraphics[width=1\textwidth]{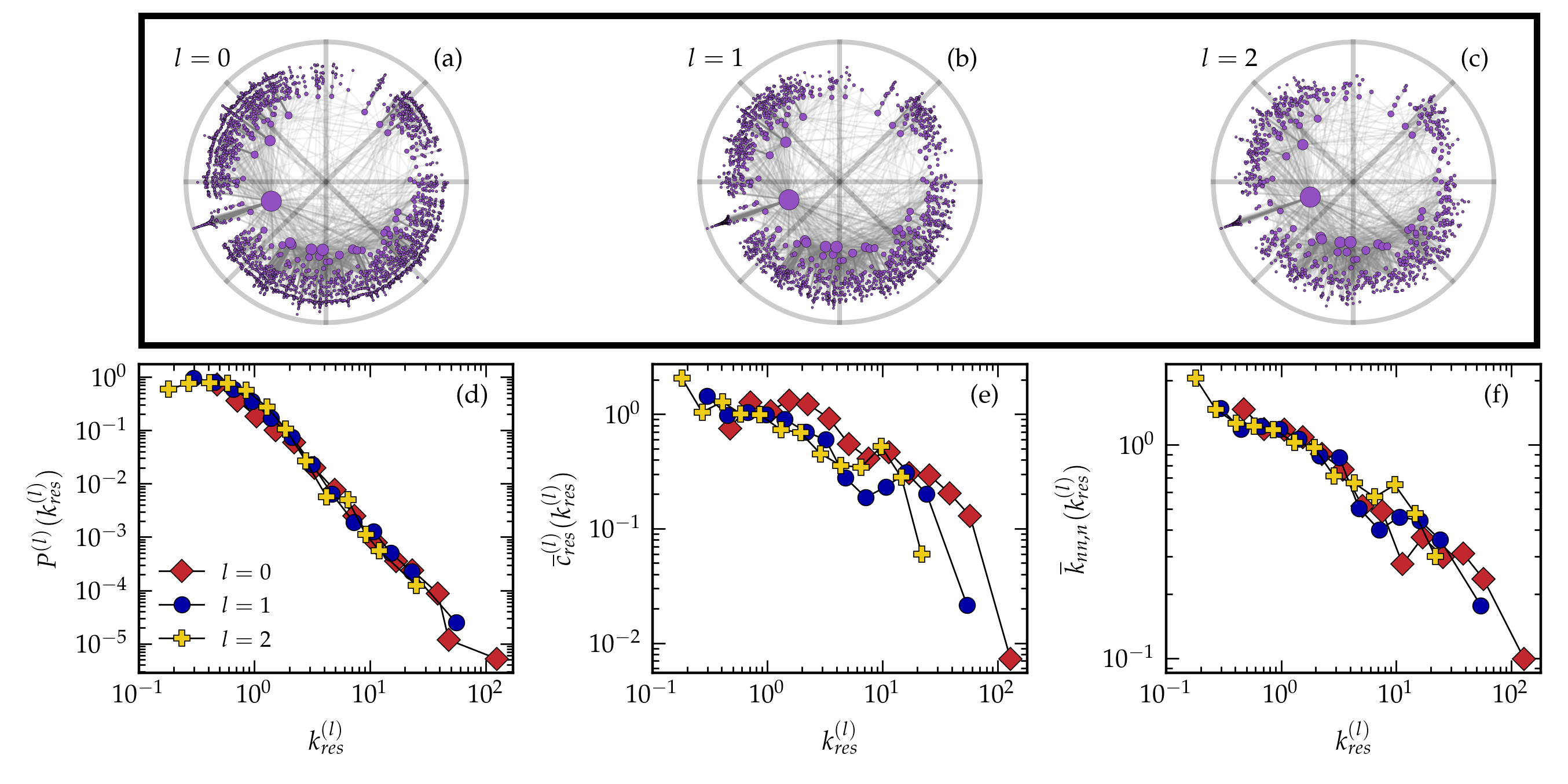}
	\vspace{-8mm}
	\caption{Summary of the results of GR for the GMP–C.elegans network. \textbf{(a-c)} Representation of the embedding for layers $l=1,2 $ and $3$ in the hyperbolic plane. The top 50\% most geometric edges are shown. The topological properties are also given: \textbf{(d)} the degree distribution, where $k_{res}^{(l)}=k^{(l)}/\langle k^{(l)}\rangle$, \textbf{(e)} the rescaled average local clustering coefficient per degree class, where $\overline c^{(l)}_{res}(k^{(l)}) = \overline c^{(l)}(k^{(l)})/\overline c^{(l)}$ and finally \textbf{(f)} the degree-degree correlations per degree class. In all cases we log-bin the degrees.  }
	\label{Sfig:GMP-Celegans_R}
\end{figure}
\begin{figure}[h]
	\centering
	\includegraphics[width=1\textwidth]{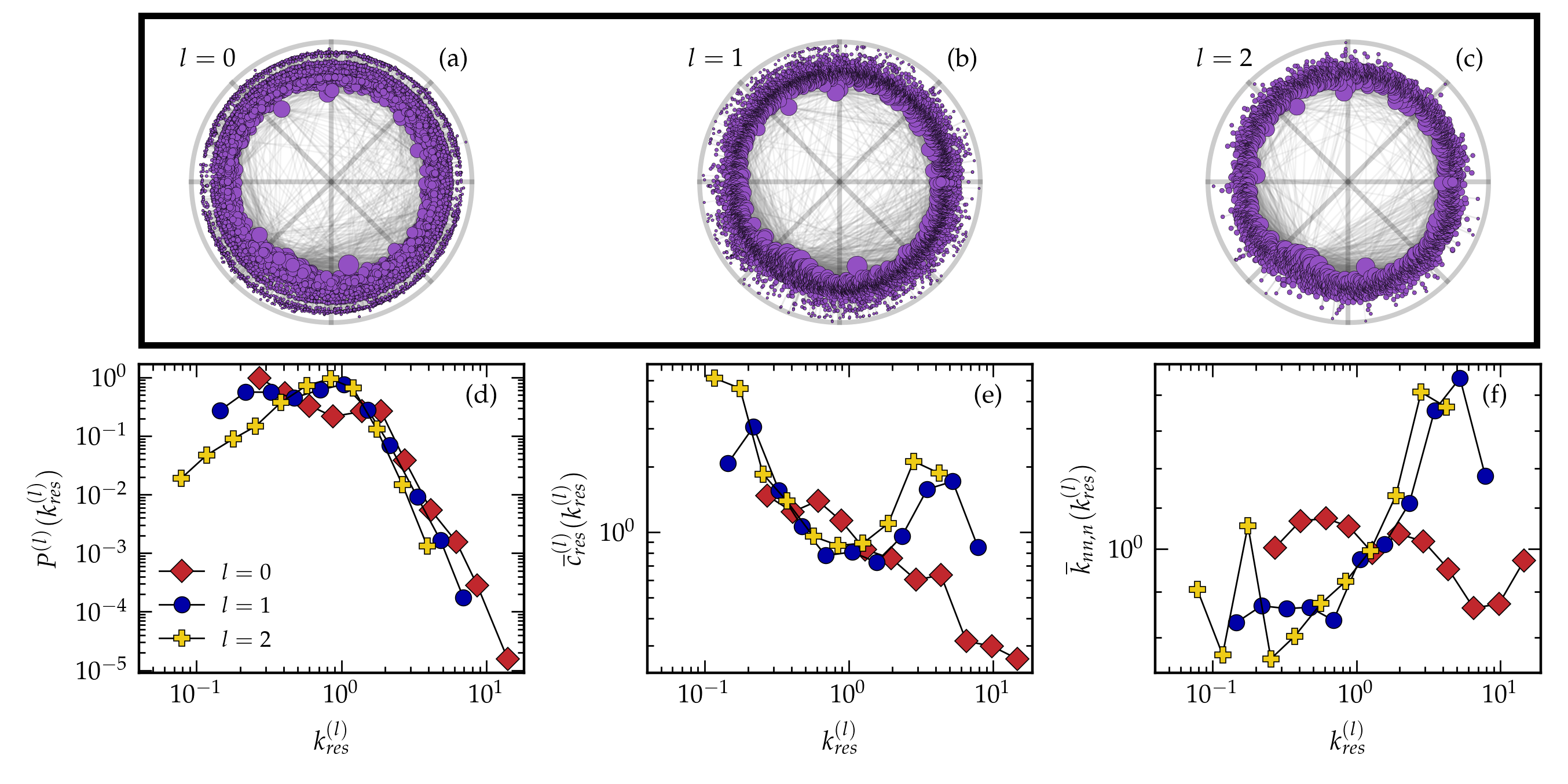}
	\vspace{-8mm}
	\caption{Summary of the results of GR for the Gnutella network. \textbf{(a-c)} Representation of the embedding for layers $l=1,2 $ and $3$ in the hyperbolic plane. The top 40\% most geometric edges are shown. The topological properties are also given: \textbf{(d)} the degree distribution, where $k_{res}^{(l)}=k^{(l)}/\langle k^{(l)}\rangle$, \textbf{(e)} the rescaled average local clustering coefficient per degree class, where $\overline c^{(l)}_{res}(k^{(l)}) = \overline c^{(l)}(k^{(l)})/\overline c^{(l)}$ and finally \textbf{(f)} the degree-degree correlations per degree class. In all cases we log-bin the degrees.  }
	\label{Sfig:Gnutella}
\end{figure}
\begin{figure}[h]
	\centering
	\includegraphics[width=1\textwidth]{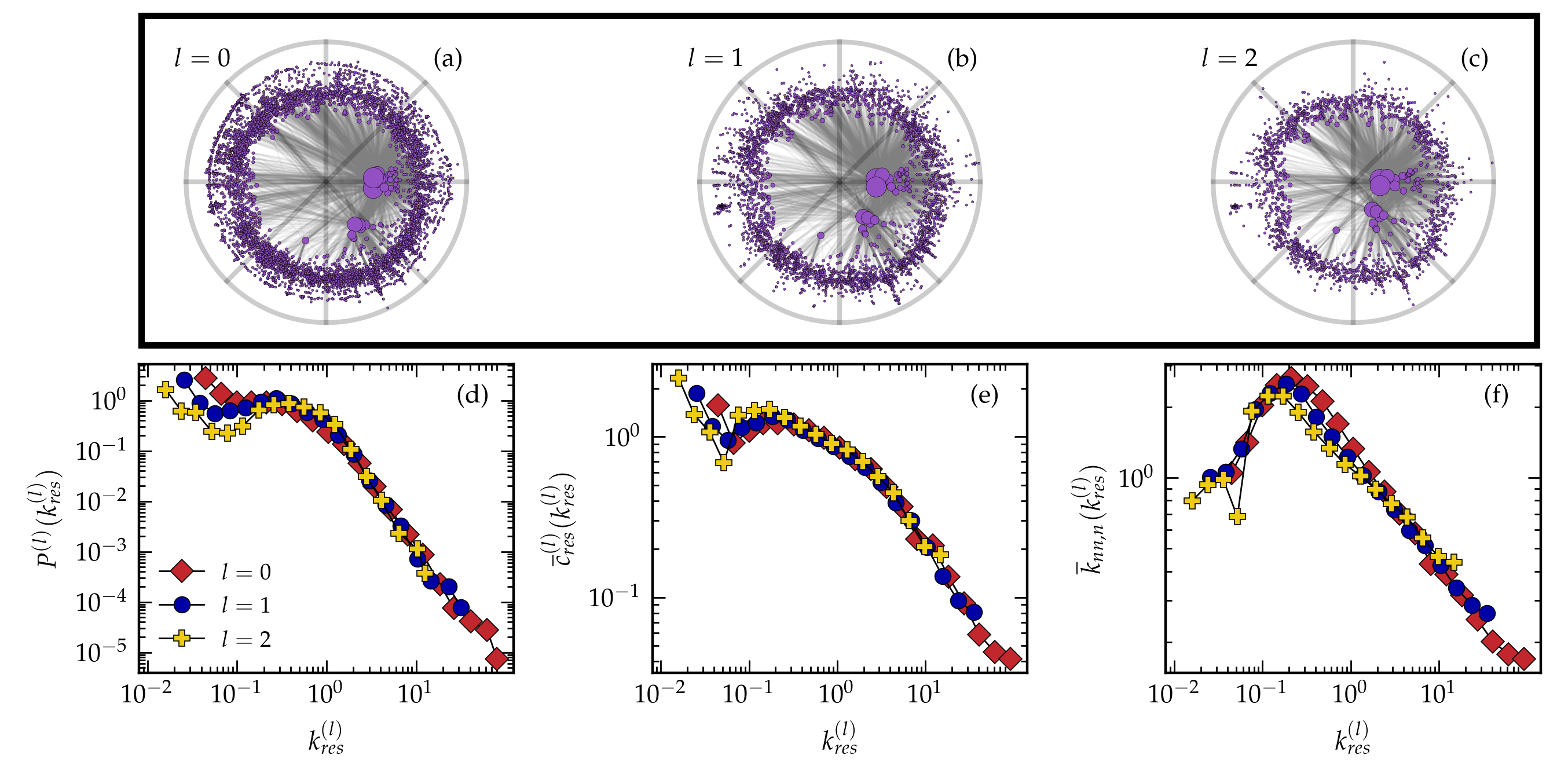}
	\vspace{-8mm}
	\caption{Summary of the results of GR for the PPI–S.cerevisiae network. \textbf{(a-c)} Representation of the embedding for layers $l=1,2 $ and $3$ in the hyperbolic plane. The top 3\% most geometric edges are shown. The topological properties are also given: \textbf{(d)} the degree distribution, where $k_{res}^{(l)}=k^{(l)}/\langle k^{(l)}\rangle$, \textbf{(e)} the rescaled average local clustering coefficient per degree class, where $\overline c^{(l)}_{res}(k^{(l)}) = \overline c^{(l)}(k^{(l)})/\overline c^{(l)}$ and finally \textbf{(f)} the degree-degree correlations per degree class. In all cases we log-bin the degrees.  }
\end{figure}
\begin{figure}[h]
	\centering
	\includegraphics[width=1\textwidth]{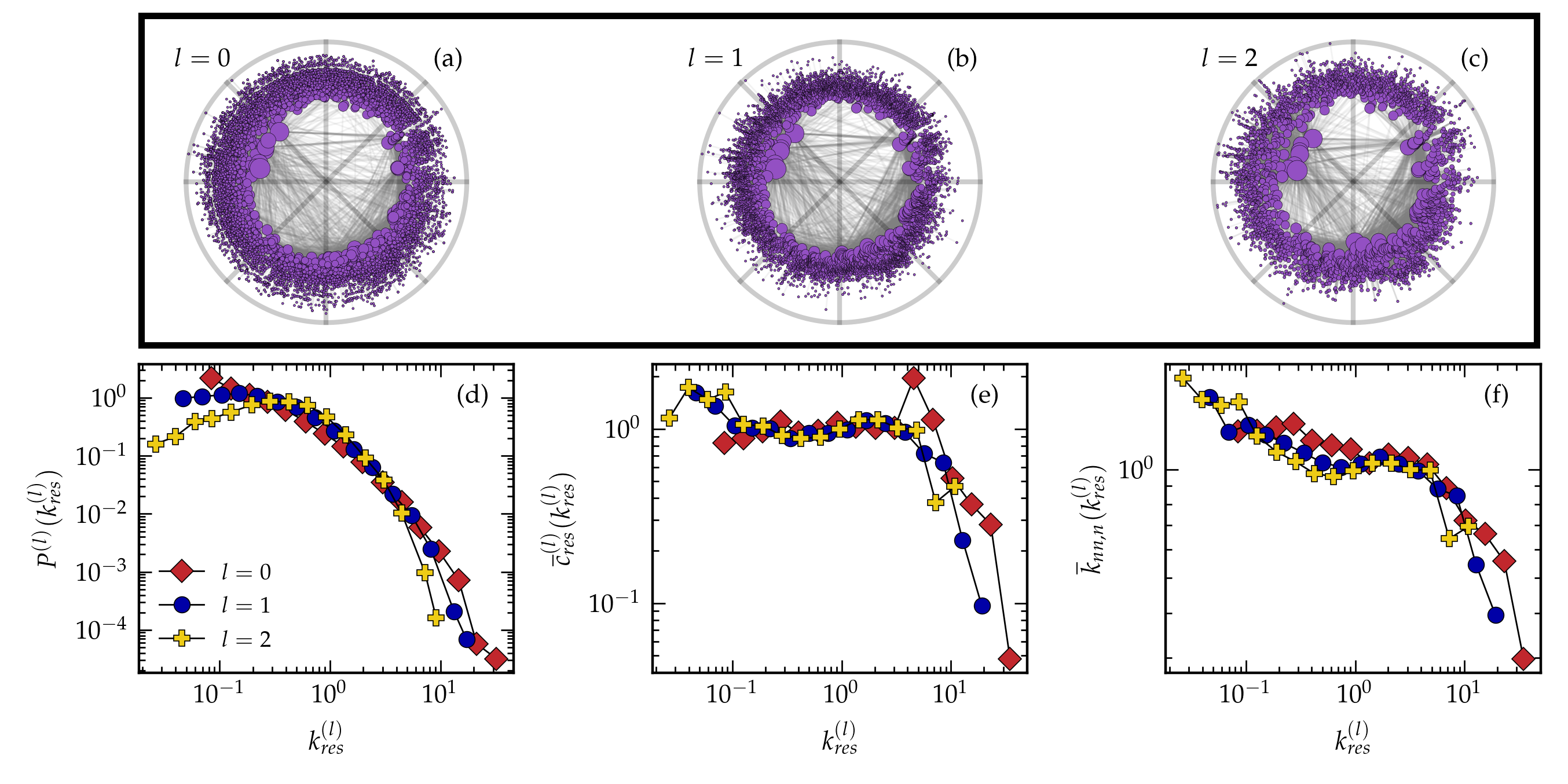}
	\vspace{-8mm}
	\caption{Summary of the results of GR for the PPI–D.melanogaster network. \textbf{(a-c)} Representation of the embedding for layers $l=1,2 $ and $3$ in the hyperbolic plane. The top 20\% most geometric edges are shown. The topological properties are also given: \textbf{(d)} the degree distribution, where $k_{res}^{(l)}=k^{(l)}/\langle k^{(l)}\rangle$, \textbf{(e)} the rescaled average local clustering coefficient per degree class, where $\overline c^{(l)}_{res}(k^{(l)}) = \overline c^{(l)}(k^{(l)})/\overline c^{(l)}$ and finally \textbf{(f)} the degree-degree correlations per degree class. In all cases we log-bin the degrees.  }
\end{figure}
\begin{figure}[h]
	\centering
	\includegraphics[width=1\textwidth]{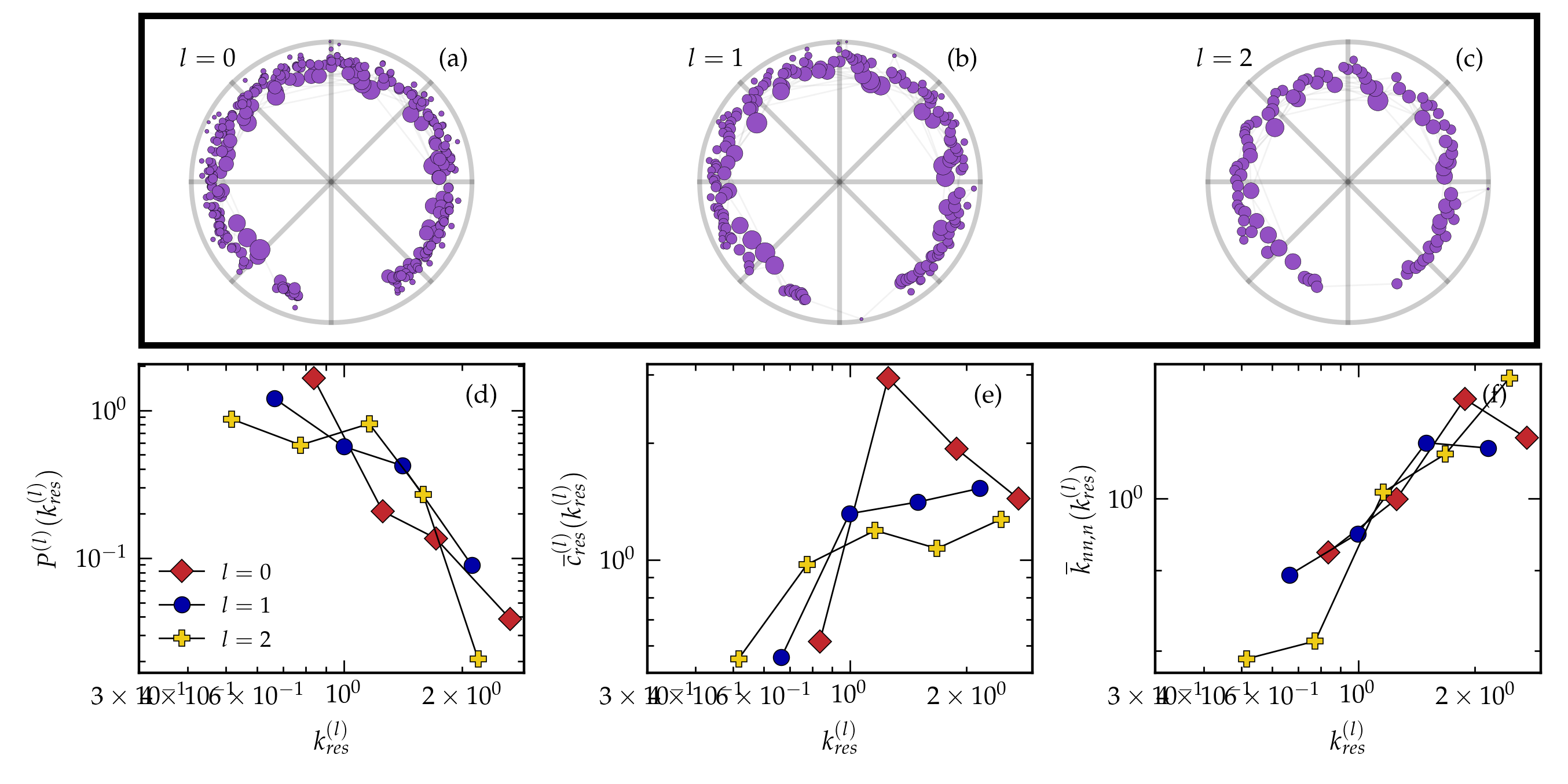}
	\vspace{-8mm}
	\caption{Summary of the results of GR for the Transport–London network. \textbf{(a-c)} Representation of the embedding for layers $l=1,2 $ and $3$ in the hyperbolic plane. The top 100\% most geometric edges are shown. The topological properties are also given: \textbf{(d)} the degree distribution, where $k_{res}^{(l)}=k^{(l)}/\langle k^{(l)}\rangle$, \textbf{(e)} the rescaled average local clustering coefficient per degree class, where $\overline c^{(l)}_{res}(k^{(l)}) = \overline c^{(l)}(k^{(l)})/\overline c^{(l)}$ and finally \textbf{(f)} the degree-degree correlations per degree class. In all cases we log-bin the degrees.  }
\end{figure}
\begin{figure}[h]
	\centering
	\includegraphics[width=1\textwidth]{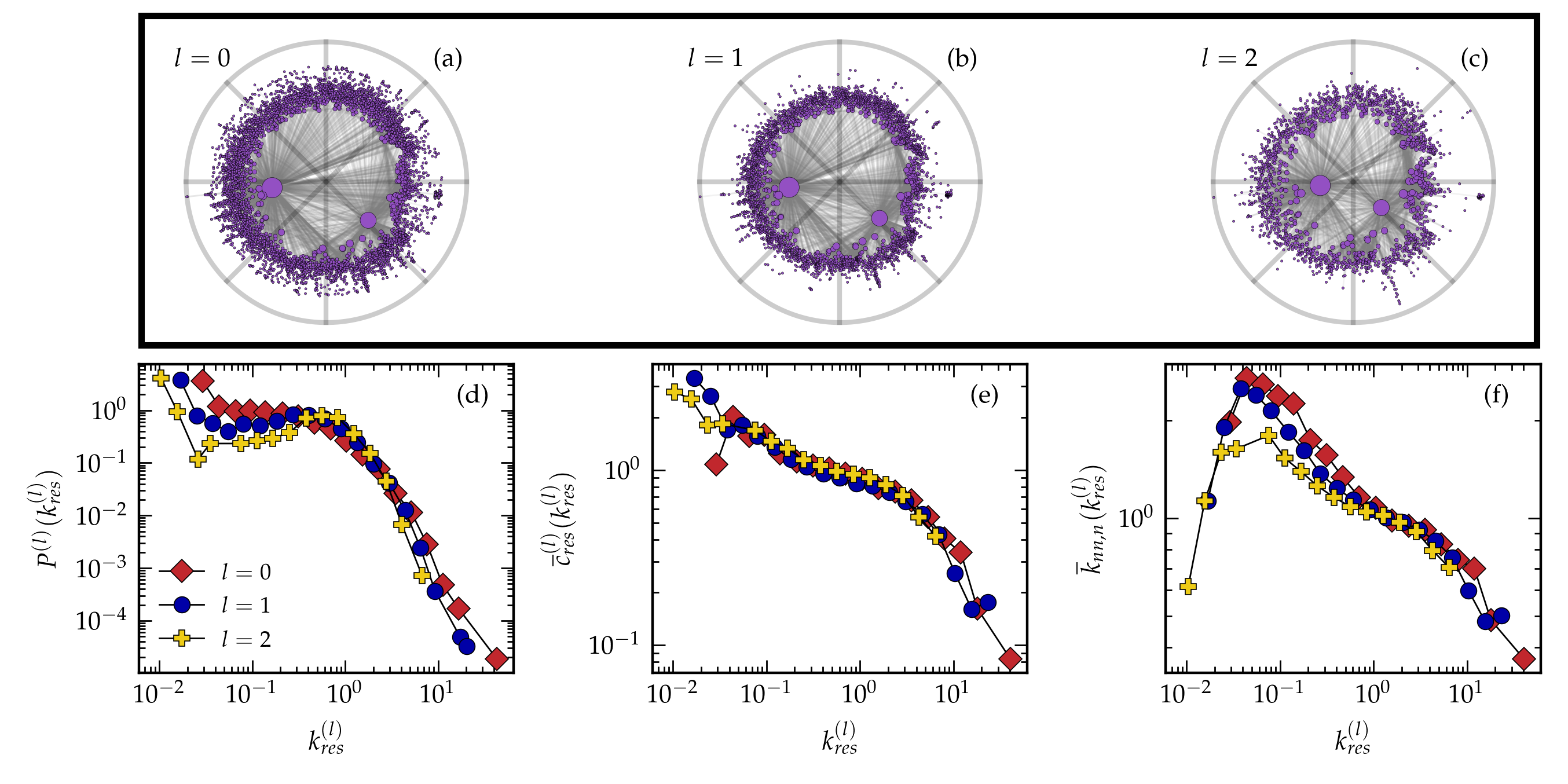}
	\vspace{-8mm}
	\caption{Summary of the results of GR for the GMP–S.cerevisiae network. \textbf{(a-c)} Representation of the embedding for layers $l=1,2 $ and $3$ in the hyperbolic plane. The top 5\% most geometric edges are shown. The topological properties are also given: \textbf{(d)} the degree distribution, where $k_{res}^{(l)}=k^{(l)}/\langle k^{(l)}\rangle$, \textbf{(e)} the rescaled average local clustering coefficient per degree class, where $\overline c^{(l)}_{res}(k^{(l)}) = \overline c^{(l)}(k^{(l)})/\overline c^{(l)}$ and finally \textbf{(f)} the degree-degree correlations per degree class. In all cases we log-bin the degrees.  }
\end{figure}
\begin{figure}[h]
	\centering
	\includegraphics[width=1\textwidth]{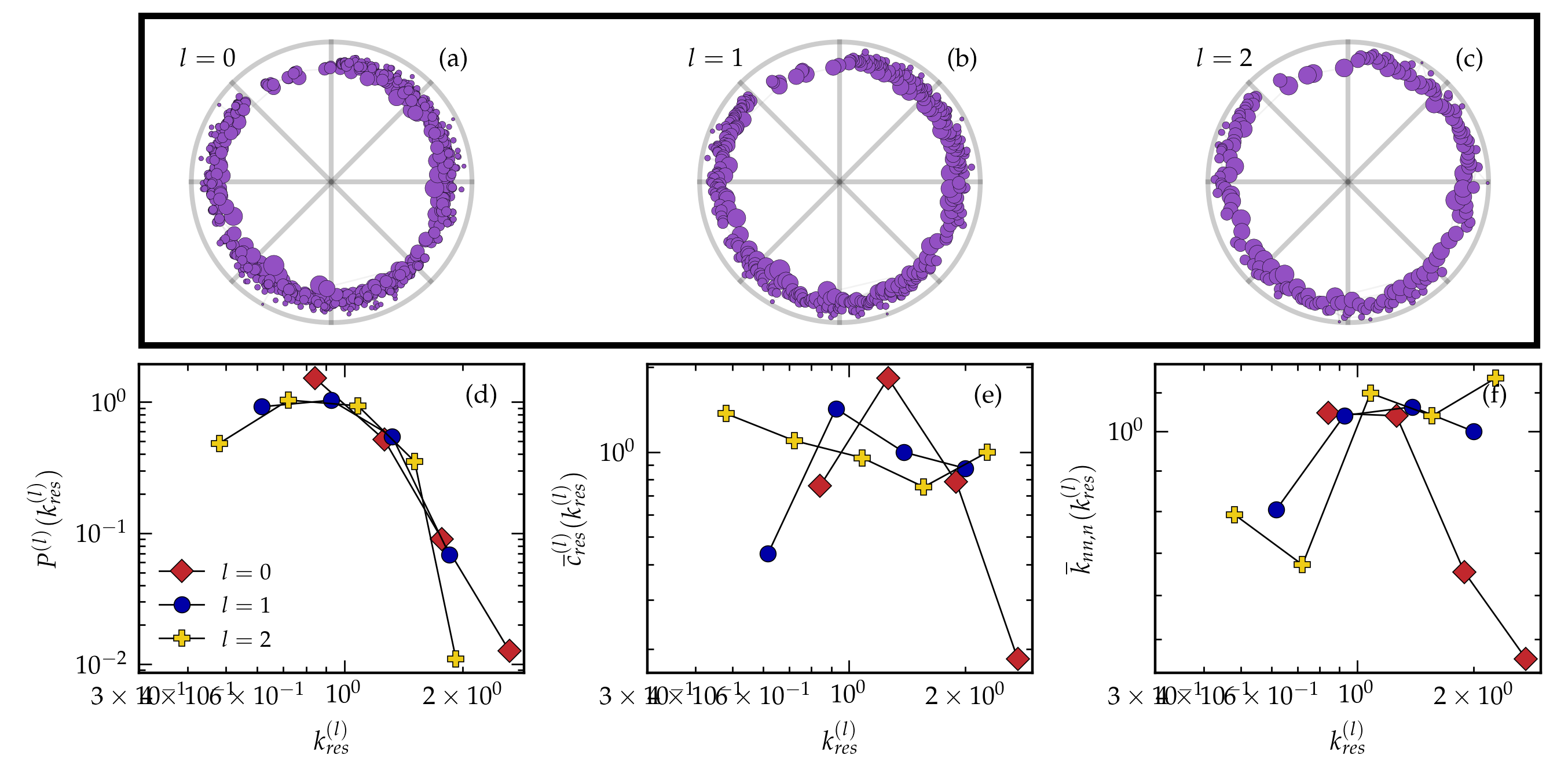}
	\vspace{-8mm}
	\caption{Summary of the results of GR for the Internet-PoP network. \textbf{(a-c)} Representation of the embedding for layers $l=1,2 $ and $3$ in the hyperbolic plane. The top 100\% most geometric edges are shown. The topological properties are also given: \textbf{(d)} the degree distribution, where $k_{res}^{(l)}=k^{(l)}/\langle k^{(l)}\rangle$, \textbf{(e)} the rescaled average local clustering coefficient per degree class, where $\overline c^{(l)}_{res}(k^{(l)}) = \overline c^{(l)}(k^{(l)})/\overline c^{(l)}$ and finally \textbf{(f)} the degree-degree correlations per degree class. In all cases we log-bin the degrees.  }
	\label{Sfig:fooweb_baywet_R}
\end{figure}
\begin{figure}[h]
	\centering
	\includegraphics[width=1\textwidth]{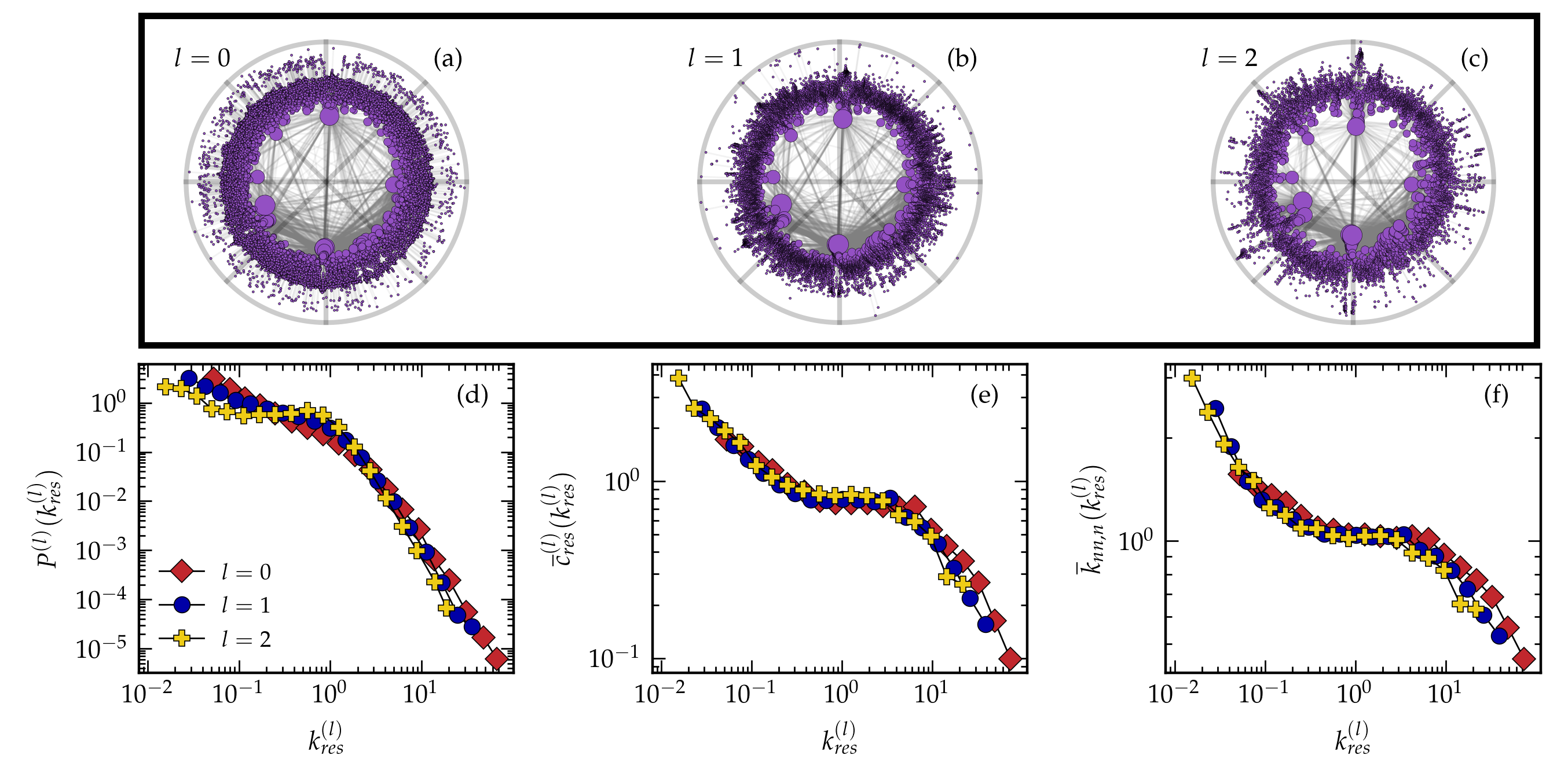}
	\vspace{-8mm}
	\caption{Summary of the results of GR for the PPI–H.sapiens network. \textbf{(a-c)} Representation of the embedding for layers $l=1,2 $ and $3$ in the hyperbolic plane. The top 5\% most geometric edges are shown. The topological properties are also given: \textbf{(d)} the degree distribution, where $k_{res}^{(l)}=k^{(l)}/\langle k^{(l)}\rangle$, \textbf{(e)} the rescaled average local clustering coefficient per degree class, where $\overline c^{(l)}_{res}(k^{(l)}) = \overline c^{(l)}(k^{(l)})/\overline c^{(l)}$ and finally \textbf{(f)} the degree-degree correlations per degree class. In all cases we log-bin the degrees.  }
\end{figure}
\begin{figure}[b]
	\centering
	\includegraphics[width=1\textwidth]{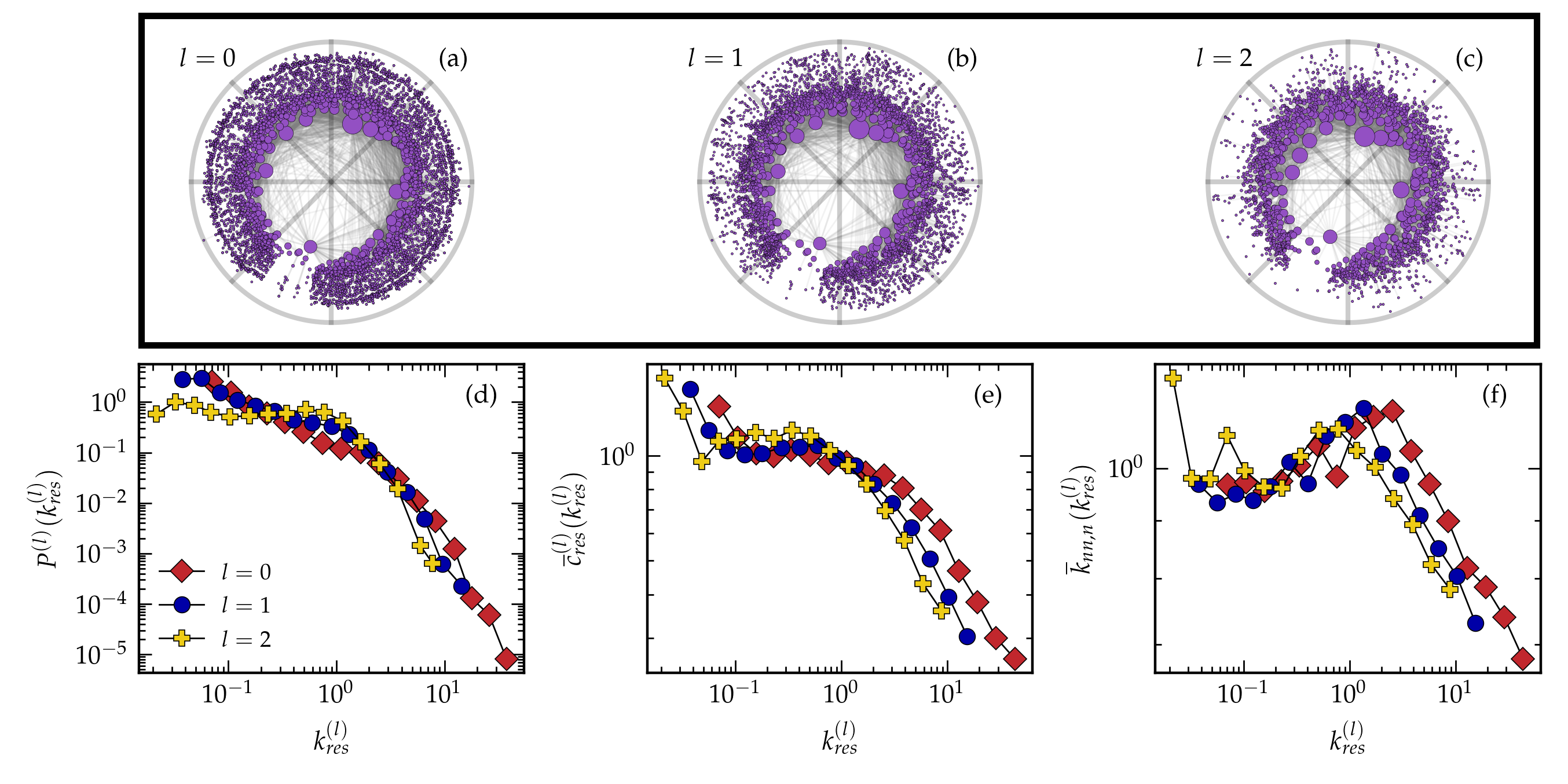}
	\vspace{-8mm}
	\caption{Summary of the results of GR for the WikiVote network. \textbf{(a-c)} Representation of the embedding for layers $l=1,2 $ and $3$ in the hyperbolic plane. The top 10\% most geometric edges are shown. The topological properties are also given: \textbf{(d)} the degree distribution, where $k_{res}^{(l)}=k^{(l)}/\langle k^{(l)}\rangle$, \textbf{(e)} the rescaled average local clustering coefficient per degree class, where $\overline c^{(l)}_{res}(k^{(l)}) = \overline c^{(l)}(k^{(l)})/\overline c^{(l)}$ and finally \textbf{(f)} the degree-degree correlations per degree class. In all cases we log-bin the degrees.  }
\end{figure}
\begin{figure}[b]
	\centering
	\includegraphics[width=1\textwidth]{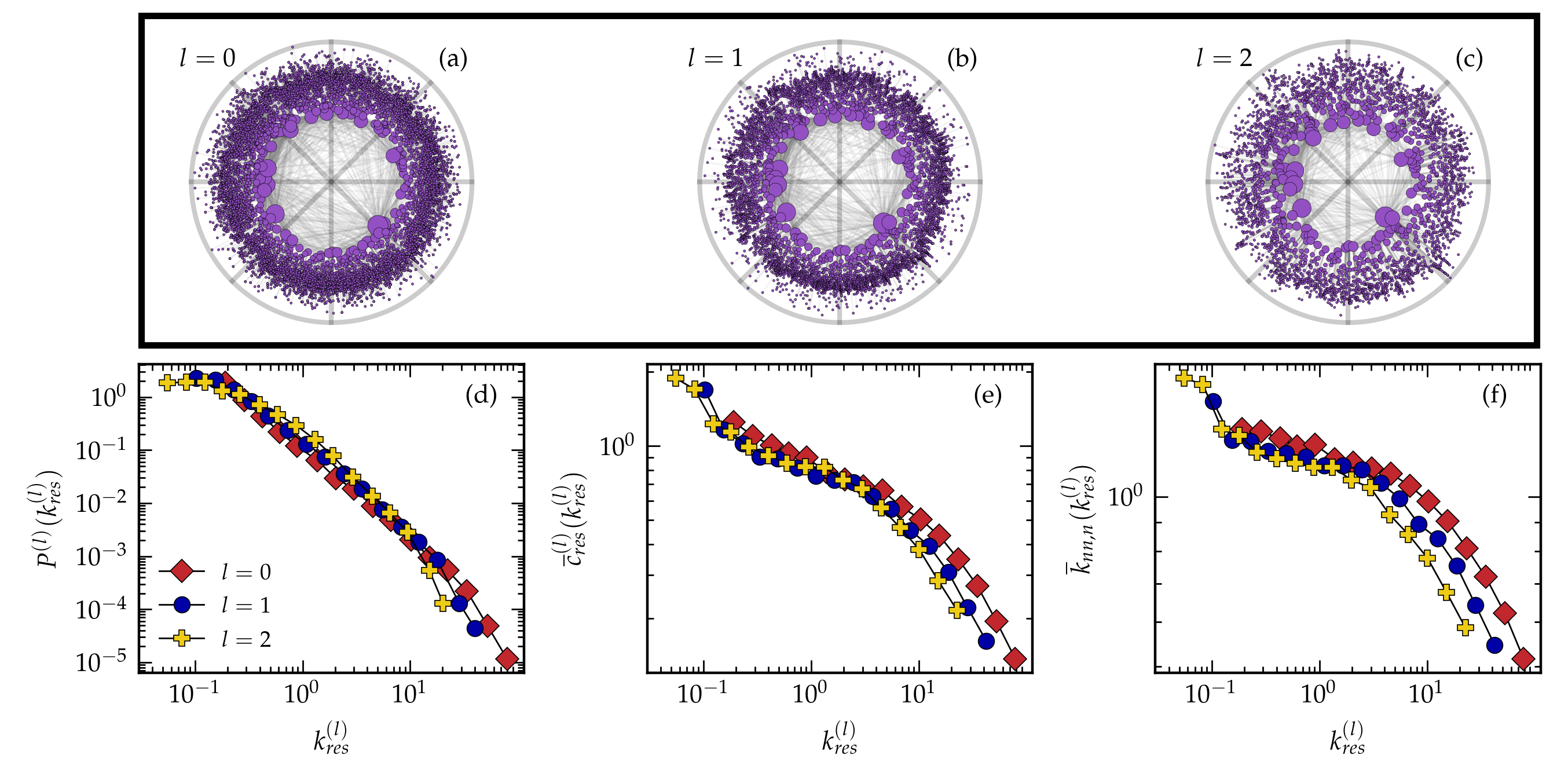}
	\vspace{-8mm}
	\caption{Summary of the results of GR for the MathOverflow network. \textbf{(a-c)} Representation of the embedding for layers $l=1,2 $ and $3$ in the hyperbolic plane. The top 5\% most geometric edges are shown. The topological properties are also given: \textbf{(d)} the degree distribution, where $k_{res}^{(l)}=k^{(l)}/\langle k^{(l)}\rangle$, \textbf{(e)} the rescaled average local clustering coefficient per degree class, where $\overline c^{(l)}_{res}(k^{(l)}) = \overline c^{(l)}(k^{(l)})/\overline c^{(l)}$ and finally \textbf{(f)} the degree-degree correlations per degree class. In all cases we log-bin the degrees.  }
\end{figure}
\clearpage
\bibliography{bibliography}

%